# The Economic Value of User-Tracking for Publishers

René Laub, Klaus M. Miller, Bernd Skiera

April 14, 2024

Rene Laub, Doctoral Candidate, Department of Marketing, Faculty of Economics and Business, Goethe University Frankfurt, Theodor-W.-Adorno-Platz 4, 60323 Frankfurt, Germany, Phone +49-69-798-34565, rlaub@wiwi.uni-frankfurt.de.

Klaus M. Miller, Assistant Professor of Marketing, HEC Paris, Rue de la Libération 1, 78350 Jouy-en-Josas, France, Hi!PARIS Chairholder, Center in Data Analytics & AI for Science, Technology, Business & Society, millerk@hec.fr.

Bernd Skiera, Professor of Electronic Commerce, Department of Marketing, Faculty of Economics and Business, Goethe University Frankfurt, Theodor-W.-Adorno-Platz 4, 60323 Frankfurt, Germany, Phone +49-69-798-34649, skiera@wiwi.uni-frankfurt.de.

This project has received funding from the European Research Council (ERC) under the European Union’s Horizon 2020 research and innovation program (grant agreement No 833714).

# The Economic Value of User Tracking for Publishers

## Abstract

Regulators and browsers increasingly restrict user tracking to protect users' privacy online. In two large-scale empirical studies, we study the economic implications for publishers relying on selling advertising space to finance their content. In our first study, we draw on 42 million ad impressions from 111 publishers covering EU desktop browsing traffic in 2016. In our second study, we use 218 million ad impressions from 10,526 publishers (i.e., apps) covering EU and US mobile in-app browsing traffic in 2023. The two studies differ in the share of trackable users (Study 1: 85%; Study 2: Apple: 17%, Android: 91%). Still, we find similar average ad impression price decreases (Study 1: 18% and Study 2: 23%) when user tracking is unavailable. More than 90% of the publishers realize lower prices when selling ad impressions for untrackable users. Publishers offering content on sports, cars, lifestyle & shopping, and news & information suffer the most. Premium publishers with high-quality edited content and strong reputations, thematic-focused (niche) publishers, and smaller publishers suffer less from the unavailability of user tracking. In contrast, non-premium publishers with non-edited or user-generated content, thematic-broad (general news) publishers, and larger publishers suffer more. The availability of a user ID generates the highest value for publishers, whereas collecting a user's browsing history, perceived as intrusive by most users, generates only a small value for publishers. These results affirm that ensuring user privacy online has substantial costs for online publishers, but those costs differ across publishers and the type of collected data. This article offers suggestions to reduce these costs.

Keywords: User Privacy, Online Advertising, Privacy Regulation, Value of Information, Tracking

## 1. Introduction

In recent years, (online) user tracking has come under increasing scrutiny. Policymakers set out to protect user privacy by restricting tracking, as exemplified by the General Data Protection Regulation (GDPR) in the European Union, the California Consumer Privacy Act (CCPA) in the United States, and China's Personal Information Protection Law (PIPL). Tech companies like Apple[1] and browsers like Mozilla Firefox[2], Google Chrome[3], and Microsoft Edge[4] reacted by introducing restrictions on third-party user tracking. As a result, user tracking is increasingly unavailable for online advertising (Choi et al. 2020).

Typically, user tracking involves collecting a user[5]'s data over time and across the websites and apps they visit, drawing on tracking technologies such as cookies, pixels, fingerprinting, device IDs, or hashed IDs (e.g., a user's e-mail address). This collection covers data in various categories, such as the ads users see and their corresponding responses, such as clicks or conversions. Firms use such data to target consumers with better content and ads and measure ad performance. Removing the ability to use data may make online advertising less attractive and thereby lower ad prices. As a result, publishers' advertising revenues are expected to shrink, leaving them with fewer resources to finance their business and content.

Industry studies have predicted a drastic decline in ad performance and, by extension, publisher revenues. Google experimentally blocked user tracking and found evidence of an average revenue loss of 52% (median: 62%; Ravichandran and Korula 2019). The UK Competition and Markets Authority (2020) produced an even higher estimate (70% loss in revenue). Academic studies, by contrast, suggest that such drastic predictions may be exaggerated: Marotta et al. (2019) found an average ad price decrease for publishers of 8%; Wang et al. (2023) indicated that the loss for publishers may be even lower (5.7%).

Granted, these academic studies have focused on individual and highly sophisticated publishers, which may be less reliant on user tracking and better able to substitute third-party with first-party data as part of a larger publisher network (as in Marotta et al. 2019). They also focused on highly vertically integrated publishers that run their own ad system, similar to closed ad ecosystems and walled gardens (which we subsequently refer to as the closed web), such as Facebook (as in Wang et al. 2023) or Google. Along these lines, privacy advocates have pointed out that some publishers may benefit from removing user tracking (e.g., by replacing behavioral targeting with contextual targeting; Ryan 2021).

---

[1] https://techcrunch.com/2021/04/26/apples-app-tracking-transparency-feature-has-arrived-heres-what-you-need-to-know
[2] https://blog.mozilla.org/en/mozilla/firefox-rolls-out-total-cookie-protection-by-default-to-all-users-worldwide
[3] https://blog.google/products/chrome/privacy-sandbox-tracking-protection
[4] https://blogs.windows.com/msedgedev/2024/03/05/new-privacy-preserving-ads-api
[5] In the legal literature, users are also referred to as data subjects.

The academic literature has yet to thoroughly vet whether all publishers in the open and ad-funded internet (which we subsequently refer to as the open web) will suffer a decrease. Such uncertainty is problematic because publishers play crucial roles in the digital landscape: They provide platforms that disseminate news, information, and opinions from varied sources, ranging from groups of professional journalists to private individuals. Publishers often provide this content free of charge, which is particularly attractive for lower-income users and enables broad political and societal participation. However, publishers cannot accurately evaluate the sustainability of the ad-supported business models that finance their content without knowledge about the effects of restricted user tracking. Furthermore, policymakers designing legislation must carefully trade off firms' ability to earn profits, which allows them to create jobs and valuable content, against users' interest in protecting their privacy.

This study seeks to offer such insights by exploring the effects of restricting user tracking across many different publishers in the open web. In contrast to prior research, which has focused on studying individual publishers, we consider differences across publishers concerning the offered ad inventory, the type of content, the thematic breadth of their content (i.e., thematic-broad or thematic-focused), the quality of the publisher (i.e., premium or non-premium), and the publisher size. The latter is especially interesting as the ad industry fears that smaller publishers may suffer disproportionately from restrictions on user tracking (Beales and Stivers 2022). Our study looks at the effect of restricted user tracking across different publishers to illuminate whether these concerns have merit.

To do so, we conduct two large-scale empirical studies using data from intermediaries in the online advertising market that allow us to observe multiple publishers, in contrast to prior research. In our first study, we draw on a sample comprising 42 million ads from a large European ad exchange. These ads emerge from 111 publishers (i.e., websites) in various topic categories, such as cars, computers and technology, finance, money and real estate, and news and information. In our second study, we observe six weeks (September - October 2023) worth of ads (about 218 million) on a demand-side platform active in the US and EU (the data also covers mobile in-app browsing traffic). The ads in our second study emerge from 10,526 publishers (i.e., apps). Together, our two studies offer a comprehensive assessment of how restrictions on user tracking affect different publishers in the open web across desktop, mobile, tablet, and app traffic.

In addition, we consider the impact of different categories of data generated by user tracking, identifying whether specific categories are more or less important for publishers. By contrast, prior empirical studies have treated restrictions on user tracking as a binary: whether information from user tracking is available or not (Marotta et al. 2022). Specifically, we consider the value of the user ID itself, the browsing history, and ad recency and frequency separately to determine how these different data categories contribute to the value of user tracking for different publishers.

Our paper thus contributes to the literature by providing novel and nuanced insights into the effects of restricting user tracking. By detailing the different effects of data categories, we compare the intrusiveness of tracking against the usefulness of the collected data via tracking for the online advertising industry (Liu et al. 2021). Such knowledge may serve as a basis for more targeted regulation (Goldfarb and Que 2023) of sensitive data categories, which are less important to advertisers and publishers but strongly threaten users' privacy (and vice versa). More nuanced regulations may better balance the interests of advertisers and publishers (i.e., generating revenues) with those of users (i.e., preserving their privacy) than blanket bans on user tracking.

By drawing on large-scale observational data, we estimate counterfactual ad prices for ad impressions with and without user tracking—thereby isolating the effect of user tracking being unavailable. We isolate the value of data generated by user tracking from user data not generated by user tracking, such as information on the device and location of a user, and publisher data, such as ad slot- and context-specific data. We confirm the robustness of our results by using augmented inverse probability weighting (AIPW), applying a two-stage Heckman (1976) correction approach, and aggregating our ad impression-level data to the publisher-level.

We find that the unavailability of user tracking leads to an 18.3% average drop in ad prices for an average publisher in the EU. This finding is a novelty in the literature, which has mainly focused on the US and China, even though the EU is under substantially more scrutiny from privacy regulators. Most publishers (>90%) suffer a price decrease, whereas a small number of publishers (<10%) obtaines higher prices from ads without user tracking. Publishers offering content on sports, cars, lifestyle & shopping, and news & information suffer the most.

We then break down this effect via a heterogeneity analysis that covers publishers' offered ad inventory, the content they provide, and their size. Without tracking, advertisers pay less for highly visible ads and prefer larger, more obtrusive ads. Premium publishers with high-quality edited content and strong reputations, thematic-focused (niche) publishers, and smaller publishers suffer less from the unavailability of user tracking.

By contrast, non-premium publishers with non-edited or user-generated content, thematic-broad (general news) publishers, and larger publishers suffer more.

Afterward, we uncover the value and importance that publishers attribute to each data category generated by user tracking. We find that the availability of a user ID generates the highest value for publishers, whereas the browsing history does not add considerable value for most publishers. Premium, thematic-broad, and larger publishers have an audience with a relatively more valuable browsing history than non-premium, thematic-focused and smaller publishers. Browsing history is most important for publishers with content targeted to students, lifestyle & shopping, and least important for publishers with games or dating content.

Finally, we complement the results of our first study with a second study, which differs in several characteristics: It focuses on device IDs within the programmatic mobile in-app ad market instead of cookies (see Vranaki and Farmer 2023) within the programmatic desktop ad market (which includes some mobile and tablet browsing data). The mobile in-app market constitutes a large share of today's online advertising, but scholars have yet to explore the impact of user tracking restrictions on its ad prices.

The two studies differ in their regulatory strictness: In our first study, regulatory strictness is relatively low under the EU e-Privacy Directive, with a share of trackable users of 85%; whereas under the second study, regulatory strictness is relatively high due to various privacy changes such as the EU's General Data Protection Regulation (GDPR) or Apple's App Tracking Transparency (ATT) framework, with a share of trackable users of 91% for Android and 17% for Apple. Despite those differences, our first and second studies produce similar results: for example, an average decrease due to the unavailability of user tracking of 18.3% (Study 1) and 23.3% (Study 2) in the EU. Our findings thus help to clarify the impact of restricted user tracking for publishers and to derive strategies that publishers can use to adjust when user tracking is available for some subpopulations of users but not others (see Acquisti 2023).

## 2. Related Literature

Our study builds on and contributes to two streams of research in information systems, marketing, and economics: the literature on (i) online advertising and targeting and (ii) online advertising and privacy.

### 2.1. Related Literature on Online Advertising and Targeting

The first stream of literature that informs our work is the literature on online advertising and targeting (Choi et al. 2020 provide a review), which distinguishes between search, social media, and display ads. In search advertising, advertisers pay a fee to search engines to be displayed next to search results (Ghose and Yang 2009, Liu et al. 2010, Zhang and Feng 2011, Xu et al. 2012, Im et al. 2016, Du et al. 2017). On social media platforms, ads are mostly shown to registered users, which allows advertisers to access a rich set of variables for targeting users (e.g., the network structure itself; Bakshy et al. 2012, Zhang et al. 2016, Kwon et al. 2017, Lee et al. 2018, So and Oh 2018). In display advertising, which includes mobile (Ghose et al. 2012), video (Xiao et al. 2024), and native (Silberstein et al. 2020), ads are shown on websites and in apps. Like those on social media, mobile ads (which comprise mobile browsing and in-app ads), offer new targeting variables that can be used to purchase impressions and may impact the value of user tracking (Ghose et al. 2019).

Our study focuses on display ads (which include video and mobile ads) rather than search or social media ads. Specifically, we focus on display ads on the open web, which reflect the bulk of online advertising revenue (IAB 2023). Here, publishers are particularly reliant on (third-party) data and presumably will be impacted more strongly by restrictions on user tracking. Publishers on the open web tend to be smaller and less sophisticated. In contrast, walled gardens (i.e., closed advertising ecosystems dominated by large platforms such as Google or Facebook) dominate search and social advertising. Such walled gardens are more reliant on first-party data, especially when users employ single-sign-on technologies (i.e., signing into the search or social media site via various devices). Policymakers consider first-party data to be less privacy infringing than third-party data and have thus focused on limiting the latter (CNIL 2020). That said, replacing third-party data with first-party data is challenging for publishers in the open web, especially smaller ones (Green 2021).

The display advertising market is two-sided (Choi et al. 2020): On the one side, advertisers buy ad impressions on publishers' websites or apps to reach potential consumers. Conversely, publishers with consumers' impressions sell ad inventory to advertisers that offer the highest valuations for those impressions. Intermediaries facilitate the match between advertisers and publishers by managing data and providing optimization tools and algorithms for serving ads[6].

Most prior works have focused on the value of user tracking for advertisers and intermediaries. Several studies have documented the effectiveness of different forms of targeting based on user tracking, such as increasing purchase intentions (Sahni 2015), click-through rates (Farahat and Bailey 2012), sales (Lewis and Reiley 2014), page views and visits (Rutz and Bucklin 2012), and online searches (Ghose and Todri-

---

[6] For a discussion of the value of user tracking for users, see e.g., Mustri et al. (2022).

Adamopoulos 2016). Johnson et al. (2020) showed that users who opt out of behavioral targeting depress ad prices by 52%. In their analytical study, Marotta et al. (2022) were among the first to differentiate user tracking data into categories: They found that advertisers prefer complete information, whereas intermediaries prefer to share only a subset of information with advertisers. However, they did not focus on publishers.

Only a few studies have focused on the value of user tracking for publishers: for example, how data brokers improve the publisher's targeting capabilities (Zhang et al. 2024) or on the value of a user's browsing history for behavioral targeting (Ghosh et al. 2015). Researchers have yet to look at the value of other data generated by user tracking, like ad recency and frequency or the mere presence of a user ID. Those forms of data would enable ad performance measures, attribution modeling, and customer journey design.

Board (2009), Chen and Stallaert (2014), Hummel and McAfee (2016), and Levin and Milgrom (2010) all used analytical models to argue that information about users' demographics or interests can increase or decrease ad prices, resulting in different levels of advertiser competition. When they access user data, advertisers can explicitly select users who match their target market; higher ad prices reflect consumers' higher expected purchase likelihood. With sufficient information about users, advertisers can create very narrow user segments. Because not all advertisers are interested in the same user segments, the demand and competition for each user's attention might decrease, with depressing effects on prices. These studies show the nuanced impact of more information on ad prices publishers receive. Overall, the conclusions of analytical studies about the value of user tracking for publishers are not straightforward.

Only a small number of papers have empirically studied the value of user tracking for publishers (see Web Appendix Table A1 for a comparison). Sun et al. (2023) found a 33% decrease in product views and an 81% decrease in gross merchandise volume when user tracking was unavailable for an e-commerce (ad publisher) platform in China. The negative impact was stronger for smaller or lower-tier advertisers. However, Sun et al. (2023) did not cover content publishers and the impact on ad impression prices.

The research that most aligns with our work is Wang et al. (2023) and Marotta et al. (2019). Wang et al. (2023) studied a large US publisher dealing with the increasing unavailability of user tracking due to the GDPR. They found that GDPR compliance led to a modest 5.7% decrease in revenue per click for the publisher. The authors also explored the heterogeneity across advertisers, finding that the GDPR hurt ads for travel and financial services the most and ads for retail and consumer products the least. They also observed that the negative effect of the GDPR was less pronounced for web pages on specific topics (e.g., sports) when the topic matched the advertised product compared to web pages on general topics (e.g., assorted news). These results suggest that thematically focused ('niche') content and contextual targeting can partially compensate for, but not reverse, the loss of user tracking for the focal publisher. However, we do not know whether this finding also holds for other niche publishers and less sophisticated publishers on the open web.

We want to emphasize that Wang et al. (2023) studied a large (among the US top 50) and highly sophisticated publisher. Their publisher observed high opt-in consent rates under GDPR, used a pay-per-click model, and focused solely on native ads, which differs from the non-native ads we consider in our study. Their publisher can use contextual and behavioral targeting for advertising. It is highly sophisticated and vertically integrated, like other publishers that operate on the closed web, such as Facebook or Google, as it runs its integrated ad system. Therefore, its ad management system vastly differs from the open web. As a result, their publisher can more easily incorporate its data in designing the auction mechanism, which starkly contrasts with many other publishers on the open web.

In their unpublished working paper, Marotta et al. (2019) compared ad prices with and without user tracking for a large multi-site US publisher and found an average ad price decrease of 8% without user tracking. Like Wang et al. (2023), this relatively low price decrease could reflect that Marotta et al. (2019) used data from a single, highly sophisticated publishing network, such as Condé Nast, which runs websites with thematically focused content (e.g., fashion, sports). That thematic focus helps advertisers reach their target markets using contextual targeting, meaning that user tracking data may be less critical. Whether that finding holds for smaller or less sophisticated publishers remains unclear.

In contrast to Wang et al. (2023) and similar to our work, Marotta et al. (2019) studied the open web. Here, publishers do not run their own ad system (i.e., they are not or less vertically integrated and therefore less able to compensate for the loss in third-party with first-party data) and instead rely on intermediaries. We complement these works by studying many open-web publishers and what value they derive from user tracking.

### 2.2. Related Literature on Online Advertising and Privacy

Our work also relates to the literature on online advertising and privacy (see Acquisti et al. 2016 and Acquisti 2023 for a review). This stream focuses on the trade-off between more precise user tracking and targeting and consumer privacy concerns. Scholars in this domain have studied the impact of privacy regulation, industry self-regulation, and other technological barriers to user tracking in online advertising.

One set of studies has looked at the impact of privacy regulation. Early work by Goldfarb and Tucker (2011) used self-reported purchase intentions to investigate the impact of opt-in user tracking restrictions within the 2002 EU e-Privacy Directive on ad effectiveness and found that purchase intentions fell by 65%. Goldberg et al. (2024) documented that GDPR, which mandates an even stricter opt-in consent, may result in a fall of up to 11.7% in page views and up to 13.2% in revenues for websites. De Matos and Adjerid (2022) studied the effectiveness of different GDPR consent campaigns: They found that opt-in consent for different data types and uses increased once GDPR consent was elicited, while also causing a decline in non-consenting users from 43% to 28%. Lefrere et al. (2024) assessed the impact of the GDPR on content publishers and documented an initial reduction in the quantity and intensity of user tracking post-GDPR. However, tracking among EU websites bounced back several months post-GDPR. In addition, they found no impact of the GDPR on new content provision and engagement.

Other studies have considered the industry self-regulation of user tracking. Johnson et al. (2020) studied the impact of opt-out consumer privacy choices in the context of the AdChoices program and found that only 0.23% of American ad impressions arise from users who opted out. Cheyre et al. (2024) evaluated the impact of Apple's ATT on app developer incentives, while Aridor and Che (2024) found that ATT significantly degraded the ability of Facebook advertisers to target advertisements based on its off-platform data.

Finally, the literature has looked at the impact of privacy-enhancing technologies, which introduce technological barriers to user tracking and targeted advertising. For example, Aseri et al. (2020) used their analytical model to demonstrate that publishers can increase their revenue by discriminating ad-intensities between regular and ad-block users. Yan et al. (2022) found that adopting an ad blocker, which usually also blocks third-party user tracking, positively affected the quantity and variety of news articles users consume.

Taken together, privacy regulation, industry self-regulation, and technological tracking barriers limit user tracking and targeting in online advertising. There are some specific subpopulations of users who cannot be tracked and targeted (e.g., users of iOS devices due to Apple ATT or EU residents due to GDPR) and some who still can (Acquisti 2023). We contribute to this stream of literature with two studies that assess the impact of tracking restrictions under varying regulatory strictness. The tracking restrictions are weaker in our first than in our second study; hence, the share of ad impressions with user tracking is higher in the former (85%) than in the latter (36%).

## 3. Institutional Background of Empirical Studies

### 3.1. Definition of User Tracking

We first detail how user tracking works to understand why restrictions on it might influence publishers, with particular emphasis on the prices that publishers earn from selling their ad inventory. *User tracking* refers to the act of collecting users' data over time through repeated identifications of each user. Identification means linking several user actions on a publisher's or third-party's property, which is usually made possible by employing a user-specific identifier. The publisher's property can take many forms, such as a website, app, or connected television. The exact means of identification depends on the precise user tracking technology (see Skiera et al. 2022). Cookies are a common tracking technology: These small text files get stored in a user's browser and include a unique user identifier, such as "168'249'342'746'836'142." Other tracking technologies exist, such as tracking pixels, digital fingerprinting, or device IDs, but they ultimately provide the same outcome: unique identifiers.

It is important to distinguish between first- and third-party cookies. First-party cookies are unique to a specific website and help improve the user experience by enabling certain functionalities, such as shopping carts. Third-party cookies, called tracking cookies, usually originate from ad tech vendors like Google Doubleclick. Thus, the user ID is specific to the combination of that ad tech vendor and user. Thus, it is possible to match users' website visits, ad impressions, and clicks on ads that appear across the affiliated websites (Ghosh et al. 2015). Such data can help advertisers target users and measure ad performance. Thus, third-party cookies likely influence the prices that publishers charge for their ad inventory, as we outline next. Furthermore, most initiatives to improve user privacy seek to limit third-party cookies, which are considered stronger infringements on privacy. Hereafter, when we refer to cookies, we mean third-party cookies.

### 3.2. Data for Online Advertising

In this section, we outline the ad-selling process and the role of data in it. We further describe the data that firms can collect and how said data might generate value for advertisers and publishers. Based on these descriptions, we clarify how and why users might consider some data more private than others, thereby highlighting the nuanced trade-offs associated with online advertisers' and users' goals (Malhotra et al. 2004, Dinev and Hart 2006).

### *3.2.1. Role of Data in the Ad Selling Process*

Online advertisers pay for the opportunity to display ads to users on publishers' websites. Many publishers sell their ad inventory online using real-time bidding (RTB; Choi et al. 2020). This sophisticated and technologically complex process includes many steps; we provide a simplified summary here and direct readers to Kosorin (2016) for a detailed description. First, users visit the publisher's website, which contains ad slots. Second, the publisher contacts an ad exchange for a real-time auction among advertisers to sell its ad slots, offering the winning advertisers the right to display their ads to specific users within an ad slot. Third, the ad exchange sends out bid requests to advertisers, sometimes through intermediaries called demand-side platforms (DSPs). The bid request can contain data specific to the user (= user data) or specific to the ad slot and its context (= publisher data), which is typically the publisher's website. If the user does not turn off tracking (e.g., via the browser or device settings), the user ID is available (as in our empirical study) and shared with the bid request. With the user ID, advertisers and other intermediaries can record bid requests and thus the user's browsing history. We refer to this collection of user and publisher data as "bid request data" (see Table 1).

*Table 1. Overview of Typical Bid Request Data in Real-Time Bidding*

| **Bid Request Data** | | | |
|---|---|---|---|
| **Data Categories** | | **Common Variables** | **Example** |
| **User data** | **Data generated by user tracking** | Identifier (ID) | User- id "123-ABC-789". |
| | | Browsing history | The user has visited www.sports.com three times already. |
| | | Ad recency | The user saw the ad "ABC" two minutes ago. |
| | | Ad frequency | The user saw the ad "ABC" already four times. |
| | **Data not generated by user tracking** | Device and software (e.g., operating system (OS), web browser) | The user is browsing the internet with a Samsung tablet, Android OS, and Firefox browser. |
| | | Location of user | The user is in Paris, France. |
| | | Date and time | The time of the user's visit is 2:00 pm on a Monday. |
| **Publisher data (Non-User data)** | **Ad slot-specific data** | Position of ad slot | The ad slot is on the top of the website. |
| | | Ad format | The ad slot has a size of 728 × 90 pixels. |
| | **Context-specific data** | Content of publisher | The ad slot is on a website offering financial news written by professional journalists, such as www.financialtimes.com. |
| | | Quality of publisher | The ad slot is on a website offering high-quality edited content, such as www.theguardian.com. |
| | | Thematic-focus of publisher | The ad slot is on a finance and business news website like uk.finance.yahoo.com. |
| | | Size of publisher | The ad slot is on the website of a small publisher, such as www.paris-normandie.fr. |

Fourth, advertisers evaluate users based on the bid request data and determine their willingness to pay (WTP). Some users may appear more attractive than others, so advertisers express a higher WTP for them. The advertisers then submit a bid to the ad exchange based on their WTP. Fifth, the ad exchange determines the winning bid and establishes the advertiser's price. Sixth, the winning advertiser displays its ad to the user on the publisher's website. Seventh, the publisher receives the payment. The price the publisher receives differs from the winning bid in a second-price auction. The advertiser with the highest bid wins the auction, but the publisher only receives the second-highest bid, while the ad exchange charges fees for this service.

As this simple illustration of the RTB process indicates, the bid request data—and thus user data gained from user tracking—can strongly influence the price paid for the ad and, by extension, the publisher's revenue.

### *3.2.2. Description of User Data*

We distinguish user data generated by user tracking and by other means (see Table 1).

The online ad industry typically uses an *identifier* (ID) to track a user's actions over time, such as whether a particular user saw a specific ad. With third-party tracking, such as cookies, firms can track users across websites. Thus, with access to the ID, an advertiser can match the display of an ad to a specific user, the user's click on an ad, or a subsequent conversion (e.g., purchase) on the advertiser's website. Thereby, the advertiser learns which ad works best for which user because users often see several of the firm's ads on different websites before clicking or converting. Advertisers want to track this ad exposure and users' reactions because such information helps them improve their advertising budget allocation decisions.

By linking the bid requests to a particular user, the advertiser also learns about the user's past website visits, called the *browsing history*. Advertisers can leverage the browsing history to create a user profile containing inferences about user demographics, habits, interests, and purchase intentions. Using this knowledge to target users with ads is a practice known as behavioral targeting.

Advertisers can calculate ad recency and frequency metrics by recording when and how often a user sees a specific ad. Ad recency is the timespan since a user last saw an ad from a specific advertiser. Ad frequency counts the number of times the user previously saw an ad from this specific advertiser. Advertisers seek an optimal level of ad frequency and ad recency, whereby they try to increase users' purchase likelihood by sparking their interest while avoiding their annoyance (Todri et al. 2020). A user who is targeted too often with the same ad may tune out (Sahni et al. 2019)[7], but leaving too much time between two impressions of the same ad could also dampen its effectiveness. With the help of user tracking, advertisers can determine effective recency and frequency capping levels.

Other data specific to the user are independent of user tracking and do not require any ID, such as the user's *device and software*. Advertisers might want to treat iPhone users differently than non-iPhone users or tablet users differently than desktop users. Software details, such as the internet browser and operating system, provide information about the user's technical setup for accessing the internet and some user characteristics. A user browsing the internet with a web browser that is not usually preinstalled, such as Firefox, might be more technically sophisticated than a user with a pre-installed browser, such as Microsoft Internet Explorer, for a long time commonly used.

Knowing a user's *location* can also influence an advertiser's WTP for ads. Users living in particular areas usually exhibit similar characteristics, such as income or education. Young single professionals generally live closer to a city center than families with children, and advertisers likely prefer to target one group over the other. This form of targeting is commonly known as geo- or location-based targeting. In addition, the *time and date* usually accompany each bid request and can potentially influence an advertiser's WTP. Some advertisers prefer to display ads to users during lunchtime rather than early morning, or on Sundays rather than Mondays.

#### ***3.2.3. Description of Publisher Data***

Publisher data refer to the environment of the ad slot for sale. They are not specific to the user and are thus independent of user tracking. Publishers maintain several ad slots on their websites, such that we can distinguish between ad slot-specific data (i.e., the exact ad slot sold) and context-specific data (all of the publisher's ad slots). Ad slot-specific data include the position of the ad slot on the publisher's website or its exact format (e.g., horizontal banner with a size of 728 × 90 pixels). Context-specific data refer to the ad context, which is usually the publisher's website for online display ads. Variables that help to describe the website include the topic of publishers' content, their quality (e.g., whether the publisher offers premium edited news content or unedited user-generated content), their thematic focus (e.g., niche versus broad content), or their size (e.g., whether the publisher is small and covering local news or large and covering international news). Advertisers can use such information to conduct contextual targeting, such as when an airline targets visitors to a travel-related website. Unlike behavioral targeting, contextual targeting does not require knowledge about the user's past behavior, such as browsing history or ad exposure. Advertisers can integrate all of these data when deciding on their response to the bid request. How useful these data are for improving online ads will likely influence advertisers' WTP, which then determines the ad prices and, by extension, publishers' revenue (Ada et al. 2022).

[7] We note that the optimal level of ads from the advertiser's perspective may differ from a user's expectation, thus leading to ad annoyance. We further note that not every advertiser fully employs these capabilities to avoid a user's ad annoyance.

## 4. Description of the First Empirical Study

### 4.1. Data and Sampling Strategy

We draw on data from a large European ad exchange that reaches 84% of users in its market. The data comprise all ads from two weeks in April 2016, resulting in 1.03 billion ad impressions. The ad exchange places a cookie on a user's browser once the user visits a publisher connected to the ad exchange. With the help of this cookie, the ad exchange, publishers, and advertisers track the user. If the user does not block tracking, the cookie has a unique user ID; otherwise, the user ID is "0". In total, 876 million ad impressions (85%) had a non-0 user ID, which we defined as ad impressions with user tracking; these impressions came from 32 million unique users. By contrast, 158 million (15%) ad impressions were not linked to any user ID, which we call ad impressions without user tracking.

To achieve computational tractability, we used a stratified sampling approach to draw a random sample of 5% of all unique user IDs (~1.6 million) and extract their ad impressions. The resulting sample consisted of 36 million ad impressions from 1.3 million users.[8] To preserve the initial distribution of ad impressions identified by cookies (85%) or not (15%), we randomly sampled 6.3 million ad impressions without user tracking. In total, our sample consisted of 41.8 million ad impressions. The data are anonymized to protect users' identities, but they clearly cover desktop and mobile browsing traffic. We observed ads from 111 publishers in various topic categories, such as cars, computers & technology, finance, money & real estate, and news & information. We provide an overview of the distribution of publishers per topic category in Table A2 in the Web Appendix. Finally, our data cover 4,563 unique advertisers.

---

[8] We excluded 300,000 users who had either unsold ads, saw public service ads, or where associated with ads outside of the EU.

### 4.2. Identification Strategy

To gauge the effect of user tracking (i.e., the presence of cookies), we accounted for a user ID that enables publishers to collect a user's data and examined its impact on the ad prices that publishers receive. In an experimental sense, the presence of a cookie should correspond to the treatment of an ad impression. However, we can never observe the price for an ad from the same user with and without a user ID, which is the fundamental problem for causal inference (Holland 1986). In an ideal setting, we would randomly assign the two states—with user tracking and without—to an ad impression and then compare the average ad price with a user ID against the average ad price without a user ID. With this setting unavailable, we instead drew on large-scale observational data to estimate the counterfactual ad prices for ad impressions with and without user tracking.

A potential challenge here is the non-randomness of the presence of the user ID. Similar to Marotta et al. (2019), our data refer to 2016, before Apple introduced its App Tracking Transparency update (in 2021) and Mozilla introduced the Enhanced Tracking Protection in Firefox (in 2019). In this context, user tracking depends on a user's deliberate configuration of a particular web browser, use of privacy management apps, or a preset default (browser) configuration. If a user's device blocks tracking, there is no cookie and thus no user tracking or user ID. This deliberate configuration raises concerns about selection bias: Ads with user tracking (treatment group) might come from systematically different users than those without user tracking (control group). Users who actively block tracking might tend to be younger or more tech-savvy, leading them to select a more privacy-sensitive internet browser.

Such differences may influence the presence of user tracking and ad prices, making it challenging to identify the causal effect of user tracking on ad prices. These underlying differences likely influence the presence of user tracking, while being less likely to influence ad prices (which result from advertisers' WTP). Thus, advertisers need to learn about the underlying differences between user groups when forming their WTP, but our setting does not provide such knowledge directly. In a causal diagram, this scenario would mean no path links the unobserved differences across users to ad prices. Instead, advertisers must rely on heuristics and observed data to infer these differences among users, such as the conventional assumption that Apple users are wealthier than non-Apple users. Our data provider also assured us that our data was the same as seen by advertisers, which provides confidence our analysis is not threatened by selection into the treatment group.

Thus, we conducted a regression-based estimation to account for observable differences between ads from users with and without user tracking, as observed by the advertisers (Study 1a). We estimated the unobserved counterfactuals within a potential outcome framework (Rubin 1976), then considered two potential outcomes for each ad impression: namely, the expected ad price with and without user tracking. Furthermore, we calculated the difference between potential outcomes, which corresponds to the average treatment effect (ATE). We also calculated the average treatment effect on the treated (ATET), which captures the effect of disabling user tracking on ads from trackable users (treatment group). This effect best simulates the current market situation, where user tracking is increasingly restricted. Finally, to confirm the robustness of our results and address any remaining selection concerns, we drew on augmented inverse probability weighting (AIPW; Bang and Robins 2005; Glynn and Quinn 2010), applied a two-stage Heckman (1976) correction approach, aggregated our ad impression-level data to the publisher-level (Study 1b), and replicated the results of our first study in a second one (Study 2).

## 5. Results of the First Empirical Study: Ad Impression-Level Analysis

### 5.1. Distribution of Ad Impression Prices

Table 2 summarizes the distribution of the raw and log ad impression prices (in cost-per-mille [CPM] in US$) with and without user tracking. Figure B1 in the Web Appendix depicts the distribution of raw prices and log(prices) according to whether user tracking is present or not.

*Table 2. Study 1a: Distribution of Raw and Log Prices for Ad Impressions with and without User Tracking*

| (CPM) Price in US$ | Mean | Standard Deviation | Median | Min | q95 | q98 | Max | N (= Ad Impressions) |
|---|---|---|---|---|---|---|---|---|
| | | | *Panel A: Raw Price Distribution* | | | | | |
| With User Tracking | 0.691 | 0.950 | 0.524 | 0.002 | 1.586 | 2.862 | 131.622 | 35,515,448 |
| Without User Tracking | 0.274 | 0.324 | 0.113 | 0.003 | 0.954 | 1.063 | 18.097 | 6,252,515 |
| | | | *Panel B: Log Price Distribution* | | | | | |
| With User Tracking | -0.788 | 0.919 | -0.647 | -6.320 | 0.461 | 1.052 | 4.87 | 35,515,448 |
| Without User Tracking | -1.730 | 0.859 | -2.18 | -5.75 | -0.047 | 0.061 | 2.896 | 6,252,515 |

Notes: $N_{Total}$ = 41,767,963; CPM = price for 1,000 ad impressions, min = minimum, max = maximum, q95 = 95% quantile, q98 = 98% quantile.

As Table 2 shows, ad impressions with user tracking fetched an average CPM price (hereafter, price or $p$) of US$0.691, whereas ad impressions without user tracking garnered, on average, a price of US$0.274. The mean difference was US$0.417. We calculated their relative price difference as follows:

$$(1)\quad \text{Relative price difference} = \frac{mean\,(p_{without\,user\,tracking}) - mean\,(p_{with\,user\,tracking})}{mean(p_{with\,user\,tracking})}.$$

The relative price difference is (0.274 - 0.691) / 0.691 = -60%. Thus, publishers realized a much lower price under restricted user tracking. Most ads without user tracking fetched a price between US$ 0.10–0.20, whereas

user tracking led to a broader distribution of ad prices (SD = 0.950). Both distributions in Table 2 reveal a long right-hand tail of ad prices: For example, the difference between the 98% and 100% quantile was approximately US$128.76 for ads with user tracking.

## 5.2. Estimation of Potential Outcomes

### *5.2.1. Model Specification*

We calculated the value of user tracking by estimating the price $p_i$ of an ad impression *i* as a function of a user tracking indicator variable $I_i^{User\ tracking}$ that captures the presence of user tracking, as well as a set of variables that control for other bid request data (see Table 1). We specified the following regression for estimating the price $p_i$ of an ad impression *i*:

$$(2)\ \log(p_i) = \alpha + \beta\, I_i^{User\ tracking} + \Gamma_i^{Technical\ configuration} + \mathrm{Z}_i^{Location} + \Theta_i^{Time} + \gamma\, I_i^{Ad\ position\ above\ fold} + \mathrm{K}_i^{Ad\ format} + \mathrm{M}_i^{Publisher} + \mathrm{N}_i^{Advertiser} + u_i.$$

We controlled for user data not generated by user tracking with a broad set of variables. We detail their descriptive statistics in the Web Appendix Table A3. Specifically, the vector of indicator variables $\Gamma_i^{Technical\ configuration}$ accounted for the device × operating system × browser combination for each user (e.g., Smartphone × Android × Firefox or Tablet × Windows × Chrome). With a vector of indicator variables $\mathrm{Z}_i^{Location}$, we differentiated ads based on users' city. The vector $\Theta_i^{Time}$ contains indicator variables that control for the week, weekday, and hour of the day. To incorporate ad-slot-specific data, we used the indicator variable $I_i^{Ad\ position\ above\ fold}$ to denote whether the ad position was above the website fold, which represents a more prominent placement. The vector of indicator variables $\mathrm{K}_i^{Ad\ format}$ controlled for the exact format of an ad (e.g., large rectangle, long and narrow skyscraper). With a vector of publisher IDs $\mathrm{M}_i^{Publisher}$, we controlled for context-specific data and accounted for differences in the ad context that reflect the inherent nature of the publisher (e.g., a prominent international news publisher such as *Le Monde* versus the considerably smaller and locally focused *Le Havre Presse* from northern France). Finally, in addition to the bid request data, we include a vector of advertiser IDs $\mathrm{N}_i^{Advertiser}$ to control for differences across advertisers, particularly their general level of spending. The error term $u_i$ (~ iid) is clustered at the user level.

### *5.2.2. Results of the Regression Analysis*

According to the regression results in Table 3, the size of the coefficient for the value of tracking decreased, and the adjusted $R^2$ increased, when we successively controlled for differences in bid request data (Models 1.2–1.9), other user data (Models 1.2–1.4), publisher data (Models 1.5 –1.7), and advertisers (Model 1.9). This observation reveals that several covariates drive the differences between ads with and without user tracking. Our primary interest was expected prices; thus, we predict the counterfactual price for each ad impression based on the regression models in Table 3 Panel A and then present these price predictions in Table 3 Panel B.

The relative price difference (-60.9%) obtained from Model 1.1 corresponds to the relative difference in mean ad impression prices with and without user tracking if we do not control for any other covariates. Suppose we control for other user data, such as differences in the devices, time, and user location (Model 1.4). In that case, the relative price difference shrinks to -52.9%, mainly due to the increase in the price without user tracking (US$0.318). That is, the relative price difference in mean ad prices, equal to -60.9%, reflects other differences in ad prices beyond the presence of user tracking. Comparing Model 1.7 and Model 1.4 clarifies that the relative price difference becomes even smaller (-39.2%) due to our controls for differences among publisher data, which include differences in ad position, ad format, and the characteristics of the publisher itself. In short, we can attribute most of the stark drop in relative price difference between Model 1.1 and Model 1.7 to differences across ad formats.

Jointly controlling for other user data not generated by user tracking and publisher data in Model 1.8 yields a relative price difference of -41.6%. From the results of Models 1.8 and 1.9 (the latter adds controls for differences between advertisers), we can see the relative price difference decrease to -18.3%. This large drop may reflect that some advertisers mainly bid on ads for trackable users with user tracking.

Overall, we observe higher prices for trackable users; these ads are valuable to advertisers and, by extension, publishers. We also learn that other user data (which may reflect device-, time- or location-specific targeting) and publisher data (which may reflect contextual targeting) account for 19.3 percentage points (-60.9% - (-41.6%) = -19.3) of the raw difference between ads with user tracking and without user tracking. The remainder of 23.3 percentage points (-41.6% - (-18.3%) = -23.3) can be attributed to differences across advertisers. The 18.3 percentage points from our preferred Model 1.9 reflect the value of user tracking, which includes the value derived from ad targeting (behavioral targeting) and ad measurement (such as ad recency and frequency and the ability to measure individual-level clicks, which is helpful for attribution modeling).

The lower part of Table 3 (Panel B) indicates the potential outcomes of the treatment group (ads with user tracking). On average, the price estimates were higher for the treatment group than the control group in conditions with or without user tracking. We found values similar to the ATE when calculating the ATET and the respective relative price decrease.

Furthermore, we calculated an expected market effect, which reflects the average price decrease per ad impression if user tracking was banned. Therefore, we multiply the share of ad impressions with user tracking (85%) with the estimated average relative price change of -18.3% from Model 1.9. The result is -18.3% × 85% = -15.6%.

## 5.3. Robustness Checks

### *5.3.1. Influence of Outliers*

We repeated the estimation using an outlier-corrected data set to assess the robustness of our findings, given the long right-tail distribution of ad prices with user tracking (see Table 2). We trimmed the data at the 98% (95%) quantile of the ad price with user tracking; that is, we excluded all ads priced higher than US$2.98 (US$1.55), which left 40,937,021 (39,757,090) ads. Accordingly, we found slightly smaller potential outcomes for ad prices in the outlier-corrected data. The average ad price with user tracking became US$0.570 (US$0.529), whereas it was US$0.470 (US$0.441) for ads without user tracking, for an estimated price difference of US$-0.100 (US$-0.088). The relative price difference was -17.5% (-16.6%), (almost) identical to the value for the entire data (-18.3%). Thus, our estimates do not appear to be affected by outliers. Next, we applied two strategies to assess the robustness of these findings concerning selection concerns: AIPW and a Heckman selection approach.

### *5.3.2. Augmented Inverse Probability Weighting (AIPW)*

An inverse probability weighting (IPW) aims to balance the treatment and control groups. We created two groups with comparable covariate distributions, thus allowing for claims of ignorability in relation to the selection into treatment (Rosenbaum and Rubin 1983, Imbens 2004). We estimated the potential outcomes using AIPW, which entails less bias and more efficiency than regular inverse probability weighting, propensity score matching, or regular regression adjustment, even for severe confounds (Glynn and Quinn 2010). The AIPW estimation involves three steps: Estimate, for each ad impression, the treatment probability equal to the probability of the presence of user tracking (also referred to as a propensity score); perform a regression-based estimation of the prices, using the observed price covariates (similar to our main analysis); and then calculate the AIPW estimator (Equation (3)).

*Table 3. Study 1a: Regression Results and Price Predictions for Value of User Tracking at the Ad Impression-Level*

*Panel A: Linear Regression Model Estimation*

| **Dependent Variable: Log(Price)** | **Model 1.1** | **Model 1.2** | **Model 1.3** | **Model 1.4** | **Model 1.5** | **Model 1.6** | **Model 1.7** | **Model 1.8** | **Model 1.9** |
|---|---|---|---|---|---|---|---|---|---|
| User Tracking (1/0) | 0.934*** (0.000) | 0.747*** (0.001) | 0.754*** (0.023) | 0.753*** (0.023) | 0.901*** (0.018) | 0.478*** (0.034) | 0.497*** (0.055) | 0.539*** (0.067) | 0.200*** (0.042) |
| Other User Data | | | | | | | | | |
| Device × Operating System × Browser | | Yes | Yes | Yes | | | | Yes | Yes |
| Time (Week, Weekday, Hour of Day) | | | Yes | Yes | | | | Yes | Yes |
| Location of User (Continent, City) | | | | Yes | | | | Yes | Yes |
| Publisher Data | | | | | | | | | |
| Ad Position (Above Fold) | | | | | Yes | Yes | Yes | Yes | Yes |
| Ad Format | | | | | | Yes | Yes | Yes | Yes |
| Publisher ID | | | | | | | Yes | Yes | Yes |
| Control | | | | | | | | | |
| Advertiser ID | | | | | | | | | Yes |
| Adj. $R^2$ | 0.119 | 0.206 | 0.215 | 0.224 | 0.125 | 0.306 | 0.392 | 0.415 | 0.654 |
| N Ad Impressions | 41,767,963 | 41,767,963 | 41,767,963 | 41,767,963 | 41,767,963 | 41,767,963 | 41,767,963 | 41,767,963 | 41,767,963 |

*Panel B: Price Predictions (Potential Outcomes) and Treatment Effects*

**All Ad Impressions**

| | **Model 1.1** | **Model 1.2** | **Model 1.3** | **Model 1.4** | **Model 1.5** | **Model 1.6** | **Model 1.7** | **Model 1.8** | **Model 1.9** |
|---|---|---|---|---|---|---|---|---|---|
| **E(Price\| User Tracking =1)** | 0.689 $ | 0.666 $ | 0.667 $ | 0.666 $ | 0.685 $ | 0.659 $ | 0.668 $ | 0.671 $ | 0.652 $ |
| **E(Price\| User Tracking =0)** | 0.269 $ | 0.315 $ | 0.314 $ | 0.314 $ | 0.278 $ | 0.409 $ | 0.406 $ | 0.392 $ | 0.533 $ |
| **ATE (in US$)** | **-0.420 $** | **-0.351 $** | **-0.353 $** | **-0.352 $** | **-0.407 $** | **-0.253 $** | **-0.262 $** | **-0.279 $** | **-0.119 $** |
| **ATE (%) / Relative Price Difference** | **-60.9%** | **-52.7%** | **-52.9%** | **-52.9%** | **-59.4%** | **-38.4%** | **-39.2%** | **-41.6%** | **-18.3%** |
| N Ad Impressions | 41,767,963 | 41,767,963 | 41,767,963 | 41,767,963 | 41,767,963 | 41,767,963 | 41,767,963 | 41,767,963 | 41,767,963 |
| | | | | | | | | | |
| **E(Price \| User Tracking =1, Treatment =1)** | 0.689 $ | 0.675 $ | 0.675 $ | 0.675 $ | 0.689 $ | 0.691 $ | 0.697 $ | 0.696 $ | 0.703 $ |
| **E(Price \| User Tracking =0, Treatment =1)** | 0.269 $ | 0.320 $ | 0.317 $ | 0.318 $ | 0.280 $ | 0.428 $ | 0.424 $ | 0.406 $ | 0.576 $ |
| **ATET (in US$)** | **-0.420 $** | **-0.355 $** | **-0.358 $** | **-0.357 $** | **-0.409 $** | **-0.263 $** | **-0.273 $** | **-0.290 $** | **-0.127 $** |
| **ATET (%) / Relative Price Difference** | **-60.9%** | **-52.6%** | **-53.0%** | **-52.9%** | **-59.4%** | **-38.1%** | **-39.2%** | **-41.7%** | **-18.1%** |
| N Ad Impressions | 35,515,448 | 35,515,448 | 35,515,448 | 35,515,448 | 35,515,448 | 35,515,448 | 35,515,448 | 35,515,448 | 35,515,448 |

*$p < 0.1$; **$p < 0.05$; ***$p < 0.01$.

Notes: ATE = Average Treatment Effect, ATET = Average Treatment Effect on the Treated. The ATE and ATET in percentages correspond to the relative price change from Equation (1). Robust standard errors are in parentheses.

We adopted a common approach for the treatment probability estimation; that is, we used logistic regression to estimate the treatment probability $\hat{\pi}_i$ for each ad impression $i$. The estimated probabilities then served as weights for the observed ads. The ads in the treatment condition took an inverse weight of $1 / \hat{\pi}_i$; those in the control condition took an inverse weight of $1 / (1 - \hat{\pi}_i)$. Thus, rare observations in the respective groups receive larger weights than observations with a high likelihood of appearing in that group (Guo and Fraser 2015)[9].

We report the full logistic regression results in Table A4 in the Web Appendix[10], where we consider the balance in covariates between the treatment and control group, using the probabilities as inverse weights (Stuart 2010). We found values below 0.25 (Rubin 2001) when calculating the standardized mean differences between the treatment and control group of covariates, thus indicating a balanced data set (see Table A6 and Figure B2 in the Web Appendix).

Next, we used a linear regression (based on Equation 2) to estimate the potential outcomes of the prices across the treatment and control groups. By combining IPW with regression-based extrapolation, we can control for variables influencing the treatment probability and the outcome of interest. Estimators that combine weighting and regressions are also known as *double-robust.* To produce unbiased estimates, we only need to specify either the treatment probability model or the outcome model (price estimation) correctly. Table A4 in the Web Appendix lists the results of the price estimation. Recall that the data provider promised that we received the same information about the bid requests as the advertisers, which gives us confidence in our price estimation.

Finally, we calculated the AIPW estimator (Robins et al. 1994, Glynn and Quinn 2010) as follows:

$$(3)\quad \widehat{ATE_{AIPW}} = \frac{1}{n}\sum_{i=1}^{n}\left\{\left[\frac{T_i Y_i}{\hat{\pi}_i} - \frac{(1-T_i)Y_i}{(1-\hat{\pi}_i)}\right] - \frac{(T_i-\hat{\pi}_i)}{\hat{\pi}_i(1-\hat{\pi}_i)}\left[(1-\hat{\pi}_i)\hat{\mathbb{E}}(Y_i|T_i = 1; Z_i) + \hat{\pi}_i\hat{\mathbb{E}}(Y_i|T_i = 0; Z_i)\right]\right\}.$$

Table 4 provides the AIPW estimation results; Table A5 in the Web Appendix contains the results from the outcome estimation, which support our findings from the regression-based main analysis. We found a slightly higher ATE (US$-0.125) and a slightly more substantial relative price difference (-19.8%).

---

[9] This approach is closely connected to the notion of survey sampling weights (Horvitz and Thompson 1952; for a more detailed discussion of weighting methods, see Lunceford and Davidian 2004).

[10] They show, for example, that ads from users browsing with Internet Explorer are more likely to allow for user tracking. This finding seems plausible because installing a non-default browser (other than Internet Explorer) requires a certain level of technological sophistication.

However, we observed minimal absolute price difference estimates between the linear regression and the AIPW (US$0.006 = US$-0.119 - (US$-0.125)).

### *5.3.3. Heckman Selection Model*

We also assessed the robustness of our findings using a two-stage Heckman (1976) selection correction procedure, in which daily Google search volume for the topic area "HTTP cookie" provides the exogenous variable. This measure might influence the availability of a cookie with an ad impression, but does not influence the observed ad impression prices (see Web Appendix Figure B3). We hypothesize that an increased Google search volume is associated with a decreasing presence of cookies; the more people learn about tracking technologies, the more likely they are to turn off user tracking.

*Table 4. Study 1a: Robustness Checks for the Average Treatment Effect*

| | **Average Treatment Effect (ATE)<br>All Ad Impressions** | | | | |
|---|---|---|---|---|---|
| | **Linear Regression**<br>(Full Data) | **Linear Regression**<br>(98% Outlier Corrected Data) | **Linear Regression**<br>(95% Outlier Corrected Data) | **AIPW** | **Heckman** |
| **E(Price\| User Tracking =1)** | 0.652 $ | 0.570 $ | 0.529 $ | 0.632 $ | 0.652 $ |
| **E(Price\| User Tracking =0)** | 0.533 $ | 0.470 $ | 0.441 $ | 0.507 $ | 0.531 $ |
| **ATE (in US$)** | **-0.119 $** | **-0.100 $** | **-0.088 $** | **-0.125 $** | **-0.121 $** |
| **ATE (%) /<br>Relative Price Difference** | **-18.3%** | **-17.5%** | **-16.6%** | **-19.8%** | **-18.5%** |
| N Ad Impressions | 41,767,963 | 40,937,021[A] | 39,757,090[A] | 41,430,997[B] | 41,767,963 |

Notes: AIPW = Augmented Inverse Probability Weighting.
[A] To account for the potential effect of outliers, we excluded all ad impressions with a price > $2.98 (= 830,942 ad impressions, 1.98%) in the 98% outlier corrected data set, and all ad impressions with a price > $1.55 (= 2,010,873 ad impressions, 4.8%) in the 95% outlier corrected data set.
[B] In line with previous research, we excluded 336,966 (0.80%) ad impressions for the AIPW estimate with a very high ($\hat{\pi}_i > 0.999$) or very low ($\hat{\pi}_i < 0.001$) treatment probabilities because they could yield very high or very low weights.

In the first-stage regression, we followed standard procedures and used a probit regression to estimate the probability of user tracking for each ad impression $i$. Google search volume is a regressor (see Web Appendix Table A4). The results support the hypothesized negative relationship between Google search volume for the topic area "HTTP cookie" and the presence of user tracking ($\beta_{\log(Google\ Search\ Volume)} = -0.004, p < .01$). With these estimates from the first stage, we calculated the inverse Mills ratio as a correction term for the second stage. In the second stage, we applied Equation 2 to estimate the expected ad impression price for each ad impression, add the correction term, and calculate the potential outcomes again (see Web Appendix Table A5). Table 4 shows that the potential outcomes and ATE estimate are consistent with our main analysis.

Previous studies mainly leveraged US and Chinese data, so our findings offer novel insights into the EU online ad market, which is under more scrutiny from regulators. In this context, we established that the average ad price decreases by 18.3% when user tracking is restricted. Overall, our result is substantially larger than the findings of both Wang et al. (2023; i.e., a 5.7% decrease in publisher revenue per click) and Marotta et al. (2019; i.e., an 8% decrease in ad prices). Granted, those studies looked at either one highly sophisticated publisher (part of a large media conglomerate) or a highly vertically integrated publisher running its own ad system similar to other closed-web publishers (e.g., Facebook or Google). Instead, we focused on multiple publishers on the open web. The heterogeneity across these publishers is likely crucial to understanding the factors that determine the value of user tracking for publishers.

## 5.4. Heterogeneity across Publishers

### *5.4.1. Differences across Publishers*

To shed more light on the differences in the value of user tracking, we considered the distribution of the relative price differences across publishers (see Table 5). In our previous analysis, in which we assigned equal weight to each ad impression, we observed a relative price difference in the mean prices of -60% (n = 41,767,963). Suppose we calculate each publisher's mean relative price difference and take the average across all publishers' relative price differences. In that case, we observe a mean relative price difference per publisher of -26% (n = 111). That is, -60% represents the market-level mean; -26% is the publisher-level mean. According to Table 5, most publishers (> 90%) realized higher prices from ads with user tracking, but with substantial variation between publishers (SD = 26.8%). Interestingly, some publishers even obtained higher prices from ads without user tracking. Publishers offering sports (-71.4%), cars (-69.6%), lifestyle & shopping (-55.1%), and news & information (-50.9%) suffer the largest relative price difference. In comparison, publishers of more privacy-sensitive dating content (+20.4%) obtained higher prices from ads without user tracking (see Web Appendix Table A2). To understand the determinants of these differences, we considered three ways that publishers distinguish themselves: (i) their offered ad inventory, (ii) their offered content, and (iii) their size.

*Table 5. Study 1a: Distribution of Relative Price Differences across Publishers (n = 111)*

| | Mean | Standard Deviation | Median | Min | q95 | q98 | Max | N (= Publishers) |
|---|---|---|---|---|---|---|---|---|
| Relative Price Difference | -26.0% | 26.8% | -22.4% | -153% | 5.8% | 11.5% | 17.9% | 111 |

### *5.4.2. Differences in Ad Inventory*

Differences in publishers' ad inventory depend on the positioning and format of their ad slots, which can take the shape of regular rectangle banners, high skyscrapers, or wide billboards. The ad format (determined by the width and height of each ad slot) affects how obtrusive an ad seems to users; naturally, large ad formats are more obtrusive. Meanwhile, positioning is commonly delineated by whether the ad appears above or below the website 'fold'. Being above the fold means that the ad slot appears on the first browser window when the website loads, whereas below the fold means the user must scroll down to see the ad. This ad slot positioning influences visibility.

We are particularly interested in the interplay of the ad inventory and user tracking, so we investigated if the value of the ad format and ad positioning changed with the presence of user tracking. In other words, we considered how an ad slot's obtrusiveness and visibility might influence the ad price with user tracking. As the results of Model 2.1 in Table 6 reveal, ad inventory above the fold (with higher visibility) generated higher prices for publishers ($\beta_{above\ fold} = 0.039$). Yet this positive effect of visibility also depended on the presence of user tracking, according to the coefficient of the interaction effect in Model 2.2 in Table 6 ($\beta_{above\ fold} = -0.074, \beta_{above\ fold\ X\ user\ tracking} = 0.127$). Advertisers pay more for ads above the fold if user tracking is available; they pay marginally less for the same position if user tracking is unavailable. Consequently, the relative price difference for highly visible ads is -29.4% when user tracking is unavailable, whereas less visible ads suffer less with a relative price difference of -20.9% (see Web Appendix Table A7).

We also found higher prices for more obtrusive ads in Model 2.1 ($\beta_{large\ ad} = 0.131$), but in Model 2.2, higher prices for large ads decreased when user tracking was available ($\beta_{large\ ad} = 0.400$, $\beta_{large\ ad\ X\ user\ tracking} = -0.295$). Similarly, smaller ads suffered more from the unavailability of user tracking than larger, obtrusive ads—the relative price difference was -27.9% for the former, but only -5.7% for the latter (see Web Appendix Table A7). Advertisers seem to prefer obtrusive ads when they cannot access user tracking. On this basis, we propose that advertisers favor ads with higher visibility and user tracking, which implies higher intrusiveness (targeting) or obtrusiveness. This result aligns with Goldfarb and Tucker's (2011) argument that either intrusiveness or obtrusiveness works well for increasing purchase intentions. However, their combination raises detrimental privacy concerns. Focusing on large ad formats might be a potential strategy that publishers can use to overcome the loss of user tracking.

***5.4.3. Differences in Content***

Publishers also differ in the thematic focus and quality of their content. Therefore, we distinguished publishers by the thematic breadth of their content: either thematic-broad or thematic-focused. General news publishers, like The Guardian in the UK, offer thematically-broad content; however, a finance-related publisher like Yahoo! Finance UK focuses on a narrower audience interested in finance and business news. To account for these differences, we included a $Thematic - Focused_i$ indicator variable in our analysis, which captures whether an ad impression comes from a thematic-focused publisher.

Furthermore, we differentiated between premium and non-premium publishers. Premium publishers offer high-quality edited content that garners them a stronger reputation than non-premium publishers. Non-premium publishers instead tend to feature unedited or user-generated content (Yang et al. 2019). With a list similar to the ComScore Top 100 list, we asked three independent raters to categorize the publishers in our sample into premium and non-premium classes. We modeled their potential differences with an indicator variable for $Premium_i$. As the results in Table 6 (Models 3.1 and 3.2) show, premium publishers achieve higher prices for their ads than non-premium publishers. These premium publishers can tout their generally well-educated and employed audiences (and their presumably higher household incomes), which likely increases advertisers' WTP. However, the interaction between the premium variable and user tracking revealed an economically small, statistically significant negative effect ($Model$ 3.2: $\beta_{premium\ X\ user\ tracking} = -0.106$), implying that user tracking slightly decreases the average ad prices for premium publishers. Perhaps the reduced competition among advertisers for users, due to more information about the user being available, leads to narrower user segments and, thus, lower prices. This finding is also reflected in the relative price differences without user tracking (see Table A8 in the Web Appendix): Premium publishers suffered a lower relative price difference (-20.3%) than non-premium premium publishers (-28.5%).

Similarly, thematic-focused publishers achieved higher prices for their ads in all models, but the positive effect decreased when user tracking was available ($Model$ 3.2: $\beta_{thematic-focused\ X\ user\ tracking} = -0.258$). These findings are also reflected in the relative price differences without user tracking (see Table A8 in the Web Appendix): Thematic-focused publishers suffered a lower price difference (-0.9%) than thematic-broad publishers (-26.6%). The reasoning parallels the effect for premium publishers: Advertisers have a certain expectation about the audiences they can reach through a thematically focused publisher, which raises their WTP for ads. However, the availability of user tracking offers more information about users, which lowers advertisers' competition for any particular segment of users, which results in decreased ad prices.

*Table 6. Study 1a: Value of (A) Ad Obtrusiveness and Visibility, (B) Premium, Thematic Content, and Size*

| | (A) | | (B) | |
|---|---|---|---|---|
| **Dependent Variable: Log(Price)** | **Model 2.1** | **Model 2.2** | **Model 3.1** | **Model 3.2** |
| User Tracking | 0.303***<br>(0.036) | 0.258***<br>(0.032) | 0.191**<br>(0.074) | 0.244***<br>(0.056) |
| Ad Above Fold | 0.039***<br>(0.007) | -0.074**<br>(0.025) | | |
| Ad Above Fold × User Tracking | | 0.127***<br>(0.023) | | |
| Large Ad | 0.131***<br>(0.015) | 0.400***<br>(0.006) | | |
| Large Ad × User Tracking | | -0.295***<br>(0.014) | | |
| Premium | | | 0.277***<br>(0.075) | 0.548***<br>(0.005) |
| Premium × User Tracking | | | | -0.106*<br>(0.009) |
| Thematic-Focused | | | 0.253***<br>(0.016) | 0.512***<br>(0.014) |
| Thematic-Focused × User Tracking | | | | -0.258***<br>(0.021) |
| Publisher Size | | | -0.003<br>(0.035) | -0.184***<br>(0.039) |
| Publisher Size x User Tracking | | | | 0.094***<br>(0.035) |
| Publisher Topic Category | | | Yes | Yes |
| Device x Operating System x Browser | Yes | Yes | Yes | Yes |
| Time (Hour, Day, Week) | Yes | Yes | Yes | Yes |
| Location of User (City) | Yes | Yes | Yes | Yes |
| Advertiser ID | Yes | Yes | Yes | Yes |
| Adj. $R^2$ | 0.552 | 0.553 | 0.573 | 0.582 |
| N Ad Impressions | 41,767,963 | 41,767,963 | 41,767,963 | 41,767,963 |

$*p < 0.1$; $**p < 0.05$; $***p < 0.01$. Notes: Robust standard errors are in parentheses.

### *5.4.4. Differences in Size*

It is important to account for publishers' size differences. For example, Beales and Stivers (2022) suggested that smaller publishers suffer more from user tracking restrictions than larger ones. We tested this suggestion and found that large publishers realize, on average, lower prices ($\beta_{large} = -0.184$)—likely because they have more ad impressions to sell and greater supply leads to lower ad prices. However, large publishers can somewhat compensate for these lower prices when user tracking is available ($Model\ 3.2$: $\beta_{large\ X\ user\ tracking} = 0.094$). Therefore, large publishers suffer a stronger price decrease (-26.3%) than small publishers (-12.7%) when user tracking is unavailable (see Web Appendix Table A8).

Overall, our analysis of the heterogeneity across publishers shows that not all publishers are equally reliant on online tracking, which also helps explain the expectations of the online advertising industry and prior academic studies (e.g., Marotta et al. 2019, Wang et al. 2023).

### 5.5. Heterogeneity across Data Categories

We also sought to understand the value that publishers attribute to each data category generated by user tracking (see Table 1). Users generally perceive tracking to be intrusive, particularly when it involves their browsing history, and thus reject it (Hoofnagle et al. 2012; PEW Research Center 2019). Accordingly, restricting user tracking protects user privacy by limiting or banning observations of the user's browsing history. While this browsing history is useful for advertisers (e.g., when car brands want to target gearheads), its impact on ad prices is unclear because it increases only an advertiser's willingness to pay if it indicates that the user belongs to the advertiser's targeted group. Otherwise, the advertiser's willingness to pay decreases (e.g., Chen and Stallaert 2014, Levin and Milgrom 2010).

Attempts to safeguard a user's browsing history may also restrict other categories of data that advertisers find helpful (see Table 1). Without user tracking, advertisers cannot measure ad frequency or recency, nor can they analyze customer journeys. Likewise, advertisers cannot measure advertising performance, which makes it more difficult to allocate marketing spending. Therefore, we aim to establish the importance of different data categories for the overall value of user tracking. To do so, we must distinguish the different data from user tracking when estimating ad prices. We replaced the indicator variable $I_i^{User\ tracking}$ with variables for different data and then estimated the following regression model:

$$(4)\quad \text{Log}(p_i) = \alpha + \beta\, I_i^{ID\ available} + \Phi_i^{Browsing\ history} + \Pi_i^{Ad\ recency\ group} + \mathrm{P}_i^{Ad\ frequency\ group} + \Gamma_i^{Technical\ configuration} + \mathrm{Z}_i^{Location} + \Theta_i^{Time} + \theta\, I_i^{Ad\ position\ above\ fold} + \mathrm{K}_i^{Ad\ format} + \mathrm{M}_i^{Publisher} + \mathrm{N}_i^{Advertiser} + u_i.$$

The indicator variable $I_i^{ID\ available}$ captures whether an ad impression has an associated user ID, which is required for collecting all other user tracking data. Next, we measured the browsing history of a user with a set of continuous variables $\Phi_i^{Browsing\ history}$, which included a variable for the user's respective number of website visits of each topic category associated with ad impression *i* (until the time ad impression *i* is served). This variable approximates the richness of the information associated with cookies and signals the user's interests. In this way, we capture the information that advertisers can use for behavioral targeting. We observed 13 topic categories (Table 7). A user who visits three websites in the news category takes a value of 3 for the variable $News\ Website\ Visits_i$, for example. These browsing history variables took a value of 0 for ad impressions without a cookie, while the topic category of the currently visited publisher took a value of 1.

Our data set contains a continuous ad recency variable, which measures the time (in minutes) since the user last saw an ad from the same advertiser. An ad recency value of 2 means that a user saw an ad from the advertisers just 2 minutes ago. We created discrete ad recency groups: 11 groups for the recency values 0–10 minutes, 9 groups with buckets for each further increment (e.g., 11–20 minutes, 21–30 minutes, 31–40 minutes, up to 100 minutes), and 1 group for ad recency greater than 100 minutes. In total, our model featured 21 ad recency groups as a set of indicator variables $\Pi_i^{Ad\ recency\ group}$. For ad impressions without user tracking, ad recency was again 0. Analogously, we also created 21 discrete ad frequency (i.e., how often a user saw the ad previously) groups that were included as $\mathrm{P}_i^{Ad\ frequency\ group}$. Ad frequency was 0 for ad impressions without user tracking.

We present the regression results in Table 7. Car website visits had a significantly negative effect on ad prices, perhaps because, in general, male consumers are more interested in cars, and men are typically less valuable for online advertising than women. In addition, consumers rarely buy cars online. However, we found significantly positive effects for website visits associated with female users, products typically bought online, or both: namely, visits to websites related to computer and technology products, lifestyle and shopping, travel, and women's interests. This result corroborates prior research indicating larger ad prices for female users (Lambrecht and Tucker 2019). It is worth noting that we also found significantly negative effects for visits to websites with relatively sensitive or private content, like health or dating.

To determine the value of browsing history for publishers, we extended our estimation of the potential outcomes of our main analysis with four new scenarios characterized by differences in the presence of user tracking data (Table 8). We used Equation (4) to estimate, for each ad impression, the expected price in each scenario. The mean of the estimated prices for each scenario indicates the likely potential outcome. For scenario 1, all data from user tracking were equal to 0. The other scenarios started with this 0 value and successively added observed data corresponding to each scenario. In scenario 2, we only added the user ID variable. In scenario 3, we estimated the potential outcome if the user ID was available and we could observe browsing history variables with each ad impression. We proceed analogously for scenarios 4 and 5.

*Table 7. Study 1a: Impact of Data Categories on Price, Representing the Value of Data Categories*

| **Dependent Variable: Log(Price)** | **Model 4.1** |
|---|---|
| Identifier available | 0.317*** (0.010) |
| | |
| Car Website Visits[A] | -0.318*** (0.090) |
| Computer & Technology Website Visits[A] | 0.583*** (0.143) |
| Dating Website Visits[A] | -1.499*** (0.525) |
| Entertainment Website Visits[A] | 0.101 (0.072) |
| Games Website Visits[A] | -0.077** (0.037) |
| Health & Medicine Website Visits[A] | -0.677*** (0.063) |
| Lifestyle & Shopping Website Visits[A] | 1.351*** (0.386) |
| Finance & Real Estate Website Visits[A] | 0.130*** (0.039) |
| News & Information Portal Website Visits[A] | -0.239 (0.429) |
| Sports Website Visits[A] | -0.260*** (0.049) |
| Student's Interest Websites Visits[A] | 0.448*** (0.226) |
| Travel Website Visits[A] | 0.158* (0.085) |
| Women's Interest Website Visits[A] | 0.288** (0.135) |
| | |
| Ad Recency Groups[B] | Yes |
| Ad Frequency Groups[B] | Yes |
| | |
| Other User Data, Publisher Data & Controls | |
| Device x Operating System x Browser, Time (Week, Weekday, Hour of Day), Location of User (Continent, City), Ad Position (Above Fold), Ad Format, Publisher ID, Advertiser ID | Yes |
| Adj. R2 | 0.661 |
| N | 41,767,963 |

*$p < 0.1$; **$p < 0.05$; ***$p < 0.01$.
Notes: Robust standard errors are in parentheses.
[A] We include the standardized browsing history variables in the estimation.
[B] See Web Appendix Table A9 for all ad recency and ad frequency group coefficients.

The results for scenario 2 show that a user ID's availability creates the highest average value for publishers. With the ID available, advertisers and connected ad tech intermediaries can measure the effectiveness of their ads for a single user and thereby hone their targeting. This ability is attractive to all advertisers, regardless of the products they offer. The results from scenario 3 indicate that browsing histories do not add considerable value for publishers (only US$0.002 on average). We propose two partly related explanations for this finding: First, advertisers do not compete for the same users (based on browsing history), so prices only increase for the small share of users that attract many advertisers. Second, ad tech intermediaries involved in behavioral targeting may grab the bulk of the potential positive difference between the price paid by advertisers and the prices received by publishers. Anecdotal evidence from a Dutch publisher, the public broadcasting company NPO, supports this finding: NPO increased its revenue by cutting out ad tech intermediaries and their margins (Lomas 2020). Furthermore, the value of ad recency and frequency strictly decreased the price of an ad impression by, on average, US$-0.010 and US$-0.028,

respectively. Hence, advertisers expressed the highest WTP when displaying an ad to a user for the first time[11].

*Table 8. Study 1a: Scenarios and Expected Prices for Value of Data Categories*

| *Scenario Number* | *Description of Scenario* | *Expected Price (in US$)* | *Differences in Expected Prices* |
|---|---|---|---|
| Scenario 1 | **No (data** from) user tracking | 0.536 | - |
| Scenario 2 | Only **user ID** available | 0.737 | Scenario 2 – Scenario 1: 0.737 – 0.536 = 0.201 |
| Scenario 3 | **Cookie id and browsing history** available | 0.739 | Scenario 3 – Scenario 2: 0.739 – 0.737 = 0.002 |
| Scenario 4 | **Cookie id and ad recency** available | 0.727 | Scenario 4 – Scenario 2: 0.727 – 0.737 = -0.010 |
| Scenario 5 | **Cookie id and ad frequency** available | 0.709 | Scenario 5 – Scenario 2: 0.709 – 0.737 = -0.028 |

Notes: All numbers are rounded to 3 decimals. N = 35,515,448 (= all ad impressions with user tracking).

*Table 9. Study 1a: Price Differences for Value of Data Categories*

| | | **Price Differences (In US$)** | | | | | | | |
|---|---|---|---|---|---|---|---|---|---|
| **Value Added by Data Categories** | **Calculation:** | **Mean** | **Sd.** | **Quantiles** | | | | | |
| | | | | **0%** | **50%** | **75%** | **90%** | **98%** | **100%** |
| Value added by Browsing History | Scenario 3 - Scenario 2 | 0.002 | 0.007 | -2.391 | 0.002 | 0.003 | 0.004 | 0.009 | 2.131 |
| Value added by Ad Recency | Scenario 4 - Scenario 2 | -0.010 | 0.029 | -0.893 | 0.000 | 0.000 | 0.000 | 0.000 | 0.000 |
| Value added by Ad Frequency | Scenario 5 - Scenario 2 | -0.028 | 0.046 | -1.110 | 0.000 | 0.000 | 0.000 | 0.000 | 0.000 |

Next, we calculated the incremental value of the data categories for each user by determining the difference between the potential outcomes of scenarios 3–4 relative to scenario 2 and investigating the distribution of those differences. The distribution of the value added by the browsing history (Table 9) implies that browsing history can increase ad prices by up to US$2.13 and decrease them by US$2.39, which aligns with the theoretical arguments of Levin and Milgrom (2010). However, the ad price increased substantially only for a tiny share of users (~2%); thus, the average incremental value for publishers was almost zero (US$0.002). The results in Tables Table 8 and Table 9 ultimately reveal that publishers reap little value from browsing history. Advertisers benefit more from a user ID—because it enables them to measure clicks, conversions, ad recency, and frequency—which then translates into higher revenues for publishers.

---

[11] Publishers could further preserve user privacy by displaying ads randomly to users. However, such an approach would require user tracking for an advertiser to know if (and how often) a user has seen an ad.

Finally, we explored the value added by the browsing history across publisher type, content category, and size (see Web Appendix Table A10). We found that premium publishers have an audience with a relatively more valuable browsing history (mean = 0.004) than non-premium publishers (mean = 0.002). Similarly, thematic-broad publishers (e.g., general news publishers) cater to users with a more valuable browsing history (mean = 0.002) than thematic-focused 'niche' publishers (mean = -0.002). Further, large publishers (mean = 0.002) rely more on the browsing history than small publishers (mean = -0.001). Browsing history data provided the largest incremental value for publishers with content targeted toward students (mean = 0.008) and lifestyle & shopping (mean = 0.007). Both audiences tend to be more open to deals and seducible through advertising than others. In contrast, browsing history provides a negative value for publishers with games (mean = -0.019) and dating content (mean = -0.017).

## 6. Results of the First Empirical Study: Publisher-Level Analysis

### 6.1. Motivation for Publisher-Level Analysis

To address potential concerns about the self-selection of users into user tracking in our ad impression-level analysis, we repeated the main analysis at the publisher-level. This analysis takes a different tack on the endogeneity issue related to the (un)availability of user tracking, which is not exogenous from the user perspective but from the publisher's perspective. To do so, we aggregated the ad impression-level data of 41,767,963 ad impressions to 7,939 publisher instances $j$ (publisher x device x operating system x ad position x ad format), which belong to 111 publishers.

### 6.2. Description of Model

For our publisher-level analysis, we calculated the value of user tracking by estimating the price $p_j$ of ad impressions per publisher instance $j$ as a function of a user tracking indicator variable $I_j^{User\ tracking}$ that captures the presence of user tracking, as well as a set of variables that control for other bid request data (see Table 1). We specify the following regression for estimating the price $p_j$ of ad impressions per publisher instance $j$:

$$(5)\quad \log(p_j) = \alpha + \beta\, I_j^{User\ tracking} + \Gamma_j^{Technical\ configuration} + \gamma\, I_j^{Ad\ position\ above\ fold} + \mathrm{K}_j^{Ad\ format} + \mathrm{M}_j^{Publisher} + u_j.$$

We used a broad set of variables to control for user data not generated by user tracking. Specifically, the vector of indicator variables $\Gamma_j^{Technical\ configuration}$ accounts for the device × operating system combination embodied by a user. To incorporate ad slot-specific data, we relied on the indicator variable $I_j^{Ad\ position\ above\ fold}$, which denotes whether the ad position is above the website fold. The vector of indicator variables $K_j^{Ad\ format}$ controlled for the exact format of an ad. With a vector of publisher IDs $M_j^{Publisher}$, we controlled for context-specific data that reflect the inherent nature of the publisher.

*Table 10. Study 1b: Regression Results for Value of User Tracking at the Publisher-Level*

| *Panel A: Linear Regression Model Estimation* | | | | | | |
|---|---|---|---|---|---|---|
| **Dependent Variable: Log(Price)** | **Model 5.1** | **Model 5.2** | **Model 5.3** | **Model 5.4** | **Model 5.5** | **Model 5.6** |
| User Tracking (1/0) | 1.160***<br>(0.017) | 0.866***<br>(0.016) | 0.632***<br>(0.012) | 0.691***<br>(0.012) | 0.691***<br>(0.012) | 0.656***<br>(0.179) |
| Other User Data | | | | | | |
| Device × Operating System | | Yes | | | | Yes |
| Publisher Data | | | | | | |
| Ad Position (Above Fold) | | | | Yes | Yes | Yes |
| Ad Format | | | Yes | Yes | Yes | Yes |
| Publisher ID | | | | | Yes | Yes |
| Adj. $R^2$ | 0.382 | 0.612 | 0.808 | 0.808 | 0.816 | 0.917 |
| N Publisher-Instances / N Publisher / N Ad Impressions | 7,939 / 111 / 41,767,963 | 7,939 / 111 / 41,767,963 | 7,939 / 111 / 41,767,963 | 7,939 / 111 / 41,767,963 | 7,939 / 111 / 41,767,963 | 7,939 / 111 / 41,767,963 |
| *Panel B: Price Predictions and Treatment Effects* | | | | | | |
| **All Ad Impressions** | | | | | | |
| | **Model 5.1** | **Model 5.2** | **Model 5.3** | **Model 5.4** | **Model 5.5** | **Model 5.6** |
| **E(Price\| User Tracking =1)** | 0.783 | 1.190 | 1.325 | 1.324 | 1.479 | 1.489 $ |
| **E(Price\| User Tracking =0)** | 0.246 | 0.501 | 0.704 | 0.704 | 0.741 | 0.772 $ |
| **ATE (in US$)** | **-0.537** | **-0.689** | **-0.621** | **-0.620** | **-0.737 $** | **-0.717 $** |
| **ATE (%) / Relative Price Difference** | **-68.6%** | **-57.9%** | **-46.9%** | **-46.8%** | **-49.87%** | **-48.2%** |
| N Publisher-Instances / N Publisher / N Ad Impressions | 7,939 / 111 / 41,767,963 | 7,939 / 111 / 41,767,963 | 7,939 / 111 / 41,767,963 | 7,939 / 111 / 41,767,963 | 7,939 / 111 / 41,767,963 | 7,939 / 111 / 41,767,963 |

**p* < 0.1; ***p* < 0.05; ****p* < 0.01. Notes: ATE = Average Treatment Effect. The ATE in percentages corresponds to the relative price change from Equation (1). Robust standard errors are in parentheses.

## 6.3. Description of Results

According to the regression results in Table 10 (Panel A), when we successively controlled for differences in bid request data (Models 5.2–5.6), other user data (Model 5.2), and publisher data (Model 5.3–Model 5.5), the size of the coefficient for the value of user tracking decreased, but the adjusted $R^2$ increased. From Model 5.1, we obtained a relative price difference of -68.6% in mean ad prices with and without user tracking, without controlling for other covariates (see Table 10 (Panel B)). When jointly controlling for other user and publisher data in Model 5.6, the relative price difference was -48.2%. The larger price predictions for ads with (US$1.489) and without user tracking (US$0.717) drove the differences between Model 5.1 and 5.6.

### 6.4. Comparison of Results of Ad Impression-Level and Publisher-Level Analyses

As signaled by the overlapping 95% confidence interval, we arrived at statistically indifferent estimates for the value of user tracking in Model 1.8 in our ad impression-level analysis ($\beta_{User\ Tracking} = 0.539, 95\% - CI[0.408, 0.670]$) and Model 5.6 in our publisher-level analysis ($\beta_{User\ Tracking} = 0.656, 95\% - CI[0.305, 1.007]$). Similarly, the estimated relative difference of mean ads with and without user tracking of -41.6% in our ad impression-level analysis resembled that of -48.2% in our publisher-level analysis. From these results, we conclude that our ad impression-level analysis led to a lower value of user tracking than the publisher-level analysis. Thus, should the ad impression-level analysis suffer from any remaining concerns due to user self-selection, such self-selection would not lead us to overestimate the value of user tracking for publishers relative to the publisher-level analysis.

## 7. Description of the Second Empirical Study

### 7.1. Motivation for the Second Empirical Study

We conducted a second study to generalize our findings from the first study to another context. We compare the characteristics of the first and second studies in Table 11.

*Table 11. Comparison of Characteristics of First and Second Empirical Study*

| | **Study 1** | **Study 2** |
|---|---|---|
| **Data Source** | Ad exchange (2016) | Demand-side Platform (2023) |
| **Number of Ad Impressions** | 41,767,963 | 218,394,708 |
| **Share of Trackable Ad Impressions** | 85% | Apple: 17%, Android: 91% |
| **Observation Window** | 2 weeks<br>(in April 2016) | 6 weeks<br>(Mid-September until end of October 2023) |
| **Geographical Focus** | EU | EU & US |
| **Number of Publishers** | 111 | 10,526 |
| **Type of Advertising** | Display | Display |
| **Type of Devices** | Desktop, Tablet, Mobile (Browser) | Mobile (In-App) |

### 7.2. Data and Sampling Strategy

For our second study, we drew on a data set from a demand-side platform in the programmatic mobile ad market. The demand-side platform helps advertisers buy ad inventory on mobile devices by bidding on their behalf in real-time auctions on ad exchanges and supply-side platforms. For each ad impression offered for sale, the demand-side platform receives a bid request with the following features (among others): (i) the operating system of the device on which the ad impression is being offered (i.e., Apple or Android); (ii) the availability of a device ID; (iii) the country of the user (e.g., a European (EU) country or US); and (iv) the date and time of the bid request. The demand-side platform then decides on whether to submit a bid or not.

Therefore, the number of bids is lower than the number of bid requests. In the case of bidding, the demand-side platform received information about the winning price, representing the expense of displaying the ad.

We have access to information for 31,890 publisher instances (publisher x device x operating system x ad format), corresponding to 10,526 publishers (apps) and about 218 million ad impressions observed over six weeks from mid-September to the end of October 2023. We only selected publisher instances with (i) >= 100 ad impressions with user tracking and (ii) >= 100 ad impressions without user tracking. About 36% of ad impressions contained information about the device ID; we refer to these as ad impressions with user tracking. This share of impressions with device ID is higher on Android (91%) than on Apple (17%; see Web Appendix Table A11) mainly because Apple's ATT has a strict opt-in requirement. Overall, the share of ad impressions with user tracking is lower in our second study than in our first study (85%).

## 8. Results of the Second Empirical Study

### 8.1. Distribution of Ad Impression Prices

Table 12 summarizes the distribution of the raw ad prices (CPM in US$) with and without user tracking (Web Appendix Table A12 does so for the log prices). Ads with user tracking fetched an average price of US$7.817 in the EU (US$8.577 in the US), whereas ads without user tracking fetched an average price of US$5.860 (US$4.720). The mean difference was US$1.957 (US$3.857). The relative price difference was -25% in the EU (-45% in the US), so publishers face a severe price decrease if user tracking is restricted (see also Web Appendix Figures B4 and B5). Most publishers (>90%) in the EU and all publishers in the US realized higher prices from ad impressions with user tracking. However, substantial variation appeared in the relative price difference between publishers (EU: SD = US$6.904, US: SD = US$13.443). Some publishers (<10%) in the EU even obtained higher prices from ad impressions without user tracking, but none in the US.

*Table 12. Study 2: Distribution of Ad Impression Prices with and without User Tracking*

| (CPM) Price in US$ | Mean | Standard Deviation | Median | Min | q95 | q98 | Max | N ( = Ad Impressions) |
|---|---|---|---|---|---|---|---|---|
| *Panel A: EU (N = 10,433,115)* | | | | | | | | |
| With User Tracking | 7.817 | 6.904 | 6.862 | 0.040 | 18.648 | 28.770 | 86.480 | 4,915,925 |
| Without User Tracking | 5.860 | 7.579 | 4.313 | 0.034 | 19.741 | 29.001 | 120.509 | 5,517,190 |
| Relative Price Difference | -25.0% | 9.8% | -37.1% | -15.0% | 5.9% | 0.8% | 39.3% | - |
| *Panel B: United States (N = 207,961,593)* | | | | | | | | |
| With User Tracking | 8.577 | 13.443 | 1.550 | 0.087 | 35.742 | 50.107 | 224.581 | 74,483,959 |
| Without User Tracking | 4.720 | 9.843 | 0.510 | 0.059 | 21.933 | 36.712 | 224.446 | 133,477,634 |
| Relative Price Difference | -45.0% | -26.8% | -67.1% | -32.2% | -38.6% | -26.7% | -0.1% | - |

Notes: $N_{Total}$ = 218,394,708; CPM Price = price for 1,000 ad impressions, min = minimum, max = maximum, q95 = 95% quantile, q98 = 98% quantile.

## 8.2. Estimation of Potential Outcomes

### *8.2.1. Model Specification*

We estimated the value of user tracking by estimating the price $p_j$ of ad impressions per publisher instance $j$ as a function of a user tracking indicator variable $I_j^{User\ tracking}$ that captures the presence of user tracking, as well as a set of variables that control for other bid request data (see Table 1). We specified the following regression for estimating the price $p_j$ of ad impressions per publisher instance $j$:

$$(6)\ \log(p_j) = \alpha + \beta\, I_j^{User\ tracking} + \Gamma_j^{Technical\ configuration} + \mathrm{K}_j^{Ad\ format} + u_j.$$

We employed several variables to control for user data not generated by user tracking. Web Appendix Table A11 provides detailed descriptive statistics for these variables. Specifically, the vector of indicator variables $\Gamma_j^{Technical\ configuration}$ accounted for a user's device × operating system combination. We considered differences in location by separately estimating our models for the EU and the US. The vector of indicator variables $\mathrm{K}_j^{Ad\ format}$ controlled for the ad format.

### *8.2.2. Results of the Regression Analysis*

According to the regression results in Table 13 (Panel A), when we successively controlled for differences in bid request data (Models 6.2-6.3 and Models 7.2-7.3), other user data (Model 6.2 and 7.2) and publisher data (Model 6.3 and 7.3), the size of the coefficient for the value of user tracking decreases, and the adjusted $R^2$ increases. Again, we predicted the counterfactual price for each publisher-instance based on the regression models in Table 13 (Panel A); we present these price predictions in Table 13 (Panel B).

*Table 13. Study 2: Regression Results and Price Predictions for Value of User Tracking*

*Panel A: Linear Regression Model Estimation*

| | EU | | | US | | |
|---|---|---|---|---|---|---|
| **Dependent Variable: Log(Price)** | Model 6.1 | Model 6.2 | Model 6.3 | Model 7.1 | Model 7.2 | Model 7.3 |
| User Tracking (1/0) | 0.861*** (0.001) | 0.368*** (0.001) | 0.266*** (0.001) | 1.052*** (0.000) | 0.641*** (0.029) | 0.628*** (0.015) |
| Other User Data | | | | | | |
| Operating System = iOS | | -0.714*** (0.001) | 0.010 (0.001) | | -0.674*** (0.032) | 0.007*** (0.000) |
| Publisher Data | | | | | | |
| Ad Format = Interstitial | | | 2.942*** (0.001) | | | 3.040*** (0.012) |
| Ad Format = Rewarded | | | 3.462*** (0.001) | | | 3.682*** (0.018) |
| Adj. $R^2$ | 0.066 | 0.088 | 0.771 | 0.076 | 0.089 | 0.771 |
| N Publisher-Instances / N Publishers / N Ad Impressions | 3,412 / 1,225 / 10,433,115 | 3,412 / 1,225 / 10,433,115 | 3,412 / 1,225 / 10,433,115 | 28,478 / 9,301 / 207,961,593 | 28,478 / 9,301 / 207,961,593 | 28,478 / 9,301 / 207,961,593 |

*Panel B: Price Predictions and Treatment Effects*

| | EU | | | US | | |
|---|---|---|---|---|---|---|
| | Model 6.1 | Model 6.2 | Model 6.3 | Model 7.1 | Model 7.2 | Model 7.3 |
| **E(Price\| User Tracking =1)** | 16.7 $ | 13.3 $ | 8.62 $ | 12.2 $ | 9.61 $ | 9.16 $ |
| **E(Price\| User Tracking =0)** | 7.07 $ | 9.21 $ | 6.61 $ | 4.27 $ | 5.06 $ | 4.89 $ |
| **ATE (in US$)** | **-9.63 $** | **-4.09 $** | **-2.15 $** | **-7.96 $** | **-4.55 $** | **-4.27 $** |
| **ATE (%) / Relative Price Difference** | **-57.7%** | **-30.8%** | **-23.3%** | **-65.2%** | **-47.3%** | **-46.6%** |
| N Publisher-Instances / N Publishers / N Ad Impressions | 3,412 / 1,225 / 10,433,115 | 3,412 / 1,225 / 10,433,115 | 3,412 / 1,225 / 10,433,115 | 28,478 / 9,301 207,961,593 | 28,478 / 9,301 207,961,593 | 28,478 / 9,301 207,961,593 |

*$p < 0.1$; **$p < 0.05$; ***$p < 0.01$. Notes: ATE = Average Treatment Effect. The ATE in percentages corresponds to the relative price change from Equation (1). Robust standard errors are in parentheses.

When we did not control for other covariates, the price relative price difference obtained from Models 6.1 and 7.1 was -57.7% for the EU and -65.2% for the US, respectively. When we jointly controlled for other user and publisher data in Models 6.3 and 7.3, the relative price difference decreased to -23.3% and -46.6% for the EU and the US, respectively.

Furthermore, we calculated an expected market effect (which reflects the average price decrease per ad impression if user tracking was banned) by multiplying the share of ad impressions with user tracking (EU: 47%, US: 34%) with the estimated average relative price change of -23.3% for the EU (-46.6% for the US) from our preferred models (Model 6.3 and 7.3). The result was -23.3% x 47% = -11.0% for the EU (-46.6% x 34% = -15.8% for the US).

## 8.3. Heterogeneity across Publishers

Mobile app publishers' ad inventory depends on the offered format of their ad slots, such as regular-sized rectangle banner ads, interstitial ads, or rewarded ads. Interstitial ads are large, full-screen ads that cover the interface of the app. Rewarded ads are ones that a user can choose to view in exchange for an in-app reward, such as watching a video ad to get an extra life in a game, reading a news article, or accessing Wi-Fi at an airport. The ad format determines how obtrusive the ad is to users; naturally, large ad formats such as interstitial or rewarded are more obtrusive.

*Table 14. Study 2: Regression Results for Value of Ad Obtrusiveness*

| | EU | | US | |
|---|---|---|---|---|
| **Dependent Variable: Log(Price)** | **Model 8.1** | **Model 8.2** | **Model 9.1** | **Model 9.2** |
| User Tracking | 0.266*** (0.039) | 0.956*** (0.084) | 0.628*** (0.015) | 1.071*** (0.038) |
| Ad Format = Interstitial | 2.942*** (0.032) | 3.240*** (0.040) | 3.040*** (0.012) | 3.277*** (0.016) |
| Ad Format = Interstitial × User Tracking | | -0.837*** (0.066) | | -0.595*** (0.025) |
| Ad Format = Rewarded | 3.462*** (0.043) | 3.634*** (0.056) | 3.682*** (0.018) | 3.908*** (0.023) |
| Ad Format = Rewarded × User Tracking | | -0.600** (0.087) | | -0.604*** (0.037) |
| Operating System | Yes | Yes | Yes | Yes |
| Adj. $R^2$ | 0.771 | 0.781 | 0.771 | 0.776 |
| N Publisher-Instances / N Publishers / N Ad Impressions | 3,412 / 1,225 / 10,433,115 | 3,412 / 1,225 / 10,433,115 | 28,478 / 9,301 / 207,961,593 | 28,478 / 9,301 / 207,961,593 |

*$p$ < 0.1; **$p$ < 0.05; ***$p$ < 0.01. Notes: Robust standard errors are in parentheses.

We are particularly interested in the interplay of ad inventory and user tracking, so we investigated if the value of the ad format changes in the presence of user tracking (see Web Appendix Table A11 for descriptive statistics). In other words, we considered how an ad slot's obtrusiveness might influence the ad impression price with user tracking. As the results of Model 8.1 and 9.1 in Table 14 reveal, we found higher prices for more obtrusive ads in the EU ($\beta_{interstitial}$ = 2.942 and $\beta_{rewarded}$ = 3.462 and the US $\beta_{interstitial}$ = 3.040 and $\beta_{rewarded}$ = 3.682). But in Models 8.2 and 9.2, those higher prices decreased when user tracking was available in the EU ($\beta_{interstitial}$ = 3.240 and $\beta_{interstitial\,x\,user\,tracking} = -0.837$; $\beta_{rewarded}$ = 3.634 and $\beta_{rewarded\,x\,user\,tracking} = -0.600$). Similar results occurred for the US ($\beta_{interstitial}$ = 3.277 and $\beta_{interstitial\,x\,user\,tracking} = -0.595$; $\beta_{rewarded}$ = 3.908 and $\beta_{rewarded\,x\,user\,tracking} = -0.604$). Similarly, the relative price difference for obtrusive ads was -6.0% in the EU and -28.2% in the US. In contrast, non-obtrusive ads suffered more from the unavailability of user tracking, with a relative price difference of -51.6% in the EU and -57.0% in the US (see Web Appendix Table A13).

## 9. Comparison of Results of Both Empirical Studies

The estimated relative price difference of mean ad impressions with and without user tracking per publisher-instance in Model 6.3 (-23.3% in the EU; Study 2) was similar to our estimate in Model 1.9 (-18.3%; Study 1a). We found similar results between studies despite differences in regulatory strictness, which provides additional evidence for the robustness and generalizability of our findings. As further proof of our studies' validity, our result for the relative price difference in the US (-46.6%) was similar to the result of Johnson et al. (2020) for a US sample (-52%).

Lastly, we compared our first and second studies' results regarding an ad slot's obtrusiveness. We found higher prices for more obtrusive ads in both studies (Study 1a and 2), but these prices decreased when user tracking was available. Accordingly, the relative price difference for large, obtrusive ads was -6.0% in Study 2 (-5.7% in Study 1a). The results in Study 2 again confirm the results of Study 1a, showing that advertisers seem to prefer obtrusive ads when user tracking is unavailable.

## 10. Summary and Conclusions

Due to increasing privacy regulation, industry self-regulation, and the rise of privacy-enhancing technologies, user tracking is becoming increasingly unavailable; with the number of trackable users diminishing, publishers are compelled to explore their revenue options. This research investigated how publishers in the open and ad-funded internet (i.e., the open web) are affected by the unavailability of user tracking. Specifically, we explored the heterogeneous effect of restricted user tracking across many different publishers. Our research complements prior research that has focused on studying individual and highly sophisticated publishers, which may be less reliant on user tracking and better able to substitute third-party with first-party data as part of a larger publisher network (as in Marotta et al. 2019), as well as highly vertically integrated publishers running their own ad system similar to closed advertising ecosystems and walled gardens (i.e., the closed web), such as Facebook (as in Wang et al. 2023) or Google.

Across two large-scale empirical studies that covered 260 million ad impressions displayed at 10,637 publishers, we found that ad prices drop, on average, by 18%-23% in the EU and by 47% in the US when user tracking is unavailable. Considering that some users have already disabled user tracking, this value translates into an average price decrease of all ad impressions in the market by 11%-16%. Multiplying these percentages by projected digital ad spending of US$585 billion worldwide for 2023 (eMarketer 2021), we anticipate an advertising revenue decrease of US$64-94 billion. In short, the online publishing industry likely faces a severe revenue decrease if user tracking is eliminated, assuming the current market equilibrium stays unchanged. We note that this decrease could be even higher if advertisers decide not to bid for non-trackable users (i.e., move their advertising business to more efficient advertising channels) or lower if advertisers decide to bid for non-trackable users (e.g., those that can be reached via contextual targeting).

To clarify how publishers differ in their reliance on revenue from user tracking, we also estimated the relative price change of ad impression prices for each publisher if user tracking was no longer available. Most publishers in the EU (>90%) and all publishers in the US experienced price decreases. Publishers offering sports, cars, lifestyle & shopping, and news & information suffered the most, whereas publishers offering (more privacy-sensitive) dating content realized higher prices without user tracking.

Premium publishers, thematically narrow publishers, and smaller publishers already rely less on user tracking than their counterparts. One viable strategy for publishers might be to narrow their content on a thematic basis and increase contextual targeting capabilities (see IAB 2021). Granted, they should proceed with caution, considering that contextual targeting can violate users' privacy, too. We further found that advertisers prefer large, more obtrusive ad formats when no user tracking is available. That is, publishers could change the ad formats they offer to include larger and more obtrusive ads to address changing norms.

It is important to note that publishers were not harmed by the unavailability of a user's browsing history. This finding is surprising, given that public discussions and behavioral targeting research typically focus on this data, but it also seems plausible given the caveats related to behavioral targeting. In particular, the user profiles inferred for behavioral targeting are often inaccurate (Mayer and Mitchell 2012; Neumann et al. 2019). That lack of accurate targeting can backfire by reducing purchase intentions (Summers et al. 2016) or even prompting reactance among users (Tucker 2012).

In contrast, the availability of a user ID generates value for publishers. This finding suggests that while publishers may benefit less from advertisers' ability to target ads to users behaviorally, publishers benefit from the ability to conduct ad measurement (e.g., measure ad frequency, ad recency, and clicks and conversions). Data that allow for ad measurement may also be less invasive and more beneficial to users—as long as said data are used to gauge metrics like ad frequency and recency capping.

Naturally, our use of observational data entails some limitations. First, we cannot draw on a fully randomized experimental research design. Nevertheless, we tried to come as close to the 'true' causal estimates as possible by applying different econometric strategies. In addition to controlling for a large set of covariates in the regression analysis, we utilized matching (using the AIPW estimator), evaluated exogenous variation in Google Trends data (using the Heckman selection model), aggregated our ad impression-level data to the publisher-level, and conducted a second study in a different context.

Second, our study only measured the impact of unavailable user tracking on ad prices (i.e., the price effect of the unavailability of user tracking). Our study did not measure the 'quantity' side of this situation, that is, the impact of unavailable user tracking on the number of ad impressions per publisher.

Third, we could not observe how publishers and advertisers act when user tracking is unavailable. Therefore, our estimates assume that the current market equilibrium stays constant.

Fourth, although our studies differentiated between categories of user tracking data (i.e., the existence of the user ID itself, browsing history data, as well as data on ad recency and frequency), we were unable to observe other categories of data used for ad measurement (e.g., an advertiser's expected conversion rate for an ad impression). Scholars should determine the value of other categories of user tracking data that can enable ad measurement, but not ad targeting.

Fifth, our data allowed us to observe whether user tracking was present, but not what targeting an advertiser chose. When user tracking is present, ads could be shown to users based on behavioral or contextual targeting. When user tracking is absent, ads could be based on contextual targeting. In this vein, future research should contrast We, therefore, leave contrasting trackable users who received behaviorally targeted ads to trackable users who received contextually targeted ads for future research.

Despite these limitations, our findings ultimately supplement previous research about the value of user tracking for publishers on the open web. These studies focused on individual and highly sophisticated publishers, which may be less reliant on user tracking and better able to substitute third-party with first-party data as part of a large publisher network, as in Marotta et al. (2019), who find an average ad price decrease for publishers of 8% and highly vertically integrated publishers that run their own ad system similar to closed advertising ecosystems and walled gardens such as Facebook, which is similar to the publisher in Wang et al. (2023), who find a 5.7% loss for publishers.

As our results suggest, restricting user tracking in an effort to protect users' privacy can have severe negative consequences and costs for publishers. However, if users are primarily concerned about efforts to collect their browsing history, then policymakers might limit the collection of these data instead of restricting user tracking overall (Goldfarb and Que 2023). Regulations might allow ad recency and frequency capping, which are also in users' interest for a better ad experience. They also could permit user tracking that records clicks and conversions. Maintaining the availability of user IDs can help avoid substantial ad price decreases. Publishers would suffer less, with a smaller price drop, and users would still benefit from enhanced privacy. Such a compromise solution promises to respect users' privacy, help advertisers target their ads better, and enable publishers to sell online ads at higher prices. Alternatively, the online advertising industry could consider self-regulation efforts to reach a suitable balance between their own and users' interests.

# The Economic Value of User Tracking for Publishers

## Web Appendix

*Table A1. Comparison of Empirical Studies on the Value of User Tracking for Publishers*

| Study | Data | Observations | Observation Window | Observation Level | Geographical Focus | Number of Publishers | Type of Advertising | Type of Devices | Vertical Integration of Publisher | Heterogeneity across | | |
|---|---|---|---|---|---|---|---|---|---|---|---|---|
| | | | | | | | | | | Advertisers | Publishers | Users |
| Marotta et al. 2019 | Observational data from one multisite publisher in 2016 | ~ 2 Mio. | 1 week | Individual ad impression | US | 1[a] | Display (Banner) | Desktop, Mobile, Tablet | No | Yes[c] | | |
| Wang et al. 2023 | Observational data from one publisher in 2018 | ~ 4 Mio. | 10 weeks | Aggregate | US | 1 | Display (Native) | Desktop | Yes | Yes | | |
| Sun et al. 2023 | Experimental data from one e-commerce platform in 2019 | ~ 0.6 Mio. | 7 hours | Individual user | China | 1 | Product Recommen-dations | Mobile | Yes | Yes | | |
| Our Study | Observational data from two intermediaries in 2016 and 2023 | ~ 42 Mio. (Study 1)<br>~ 31,890 (Study 2) | 2 weeks (Study 1)<br>6 weeks (Study 2) | Individual ad impression, Aggregate | EU (Study 1)<br>US & EU (Study 2) | 111[b] (Study 1)<br>10,526 (Study 2) | Display (Banner, Video, Mobile) | Desktop, Mobile, Tablet | No | Yes[c] | Yes | Yes |

Notes: PPM = Pay-Per-Impression, PPC = Pay-Per-Click, PPA = Pay-Per-Action. [a]One large publisher with 60 distinct websites. [b]Publishers with 84% reach in the respective market. [c]Both papers control for differences across advertisers using advertiser-fixed effects.

*Table A2. Study 1a: Distribution of Publishers Per Topic Category and Relative Price Differences across Publishers*

| **Publisher Topic Category** | **Share of Publishers (n = 111)** | **Avg. Price (CPM) with User Tracking** | **Avg. Price (CPM) without User Tracking** | **Absolute Price Difference** | **Relative Price Difference** |
|---|---|---|---|---|---|
| Health | 20% | 0.967 $ | 0.703 $ | -0.263 $ | -27.2% |
| News & Information | 19% | 0.668 $ | 0.328 $ | -0.340 $ | -50.9% |
| Sports | 16% | 0.562 $ | 0.161 $ | -0.401 $ | -71.4% |
| Lifestyle & Shopping | 11% | 1.539 $ | 0.691 $ | -0.848 $ | -55.1% |
| Women's Interests | 7% | 1.002 $ | 0.559 $ | -0.443 $ | -44.2% |
| Student's Interests | 6% | 0.821 $ | 0.697 $ | -0.124 $ | -15.1% |
| Games | 5% | 0.873 $ | 0.849 $ | -0.024 $ | -2.7% |
| Cars | 4% | 1.106 $ | 0.337 $ | -0.769 $ | -69.6% |
| Travel | 4% | 0.843 $ | 0.566 $ | -0.277 $ | -32.8% |
| Entertainment | 3% | 1.151 $ | 0.633 $ | -0.519 $ | -45.1% |
| Computers & Technology | 2% | 0.945 $ | 0.636 $ | -0.308 $ | -32.6% |
| Dating | 2% | 0.583 $ | 0.704 $ | 0.119 $ | 20.4% |
| Money, Finance & Real Estate | 2% | 1.054 $ | 0.673 $ | -0.381 $ | -36.1% |

*Table A3. Study 1a: Descriptive Statistics*

| Variable | | All Ad Impressions | | Ad Impressions with User Tracking | | Ad Impressions without User Tracking | |
|---|---|---|---|---|---|---|---|
| | | Number of Ad Impressions (% of total) | Avg. Price (CPM) | Number of Ad Impressions | Avg. Price (CPM) | Number of Ad Impressions. | Avg. Price (CPM) |
| **Device** | Desktop | 34,543,967 (82.70%) | 0.678 $ | 31,910,002 | 0.705 $ | 2,633,965 | 0.349 $ |
| | Smartphone | 2,679,746 (6.42%) | 0.180 $ | 1,179,826 | 0.252 $ | 1,499,920 | 0.123 $ |
| | Tablet | 1,990,033 (4.76%) | 0.682 $ | 1,097,181 | 0.807 $ | 892,852 | 0.528 $ |
| | Unknown | 2,554,217 (6.12%) | 0.385 $ | 1,328,439 | 0.637 $ | 1,225,778 | 0.111 $ |
| **Operating System** | Android | 5,116,678 (12.25%) | 0.292 $ | 2,601,298 | 0.478 $ | 2,515,380 | 0.099 $ |
| | Apple Macintosh | 1,606,277 (3.85%) | 0.548 $ | 662,930 | 0.808 $ | 943,347 | 0.365 $ |
| | Apple iOS | 1,543,166 (3.69%) | 0.604 $ | 500,762 | 0.825 $ | 1,042,404 | 0.498 $ |
| | BlackBerry OS | 13,417 (0.03%) | 0.336 $ | 13,218 | 0.339 $ | 199 | 0.191 $ |
| | Linux | 109,864 (0.26%) | 0.676 $ | 100,543 | 0.704 $ | 9,321 | 0.378 $ |
| | Microsoft Windows | 32,872,175 (78.70%) | 0.684 $ | 31,188,951 | 0.703 $ | 1,683,224 | 0.340 $ |
| | Symbian OS | 513 (0.00%) | 0.122 $ | 505 | 0.123 $ | 8 | 0.113 $ |
| | Unknown | 505,873 (1.21%) | 0.725 $ | 447,241 | 0.769 $ | 58,632 | 0.392 $ |
| **Browser** | Android | 782,883 (1.87%) | 0.386 $ | 774,737 | 0.386 $ | 8,146 | 0.455 $ |
| | Chrome | 9,558,233 (22.88%) | 0.515 $ | 6,776,586 | 0.677 $ | 2,781,647 | 0.121 $ |
| | Firefox | 16,868,093 (40.39%) | 0.679 $ | 15,797,090 | 0.701 $ | 1,071,003 | 0.351 $ |
| | Internet Explorer | 9.102,834 (21.79%) | 0.677 $ | 8,845,639 | 0.687 $ | 257,195 | 0.319 $ |
| | Opera | 447,038 (1.07%) | 0.654 $ | 384,754 | 0.707 $ | 62,284 | 0.323 $ |
| | Safari | 2,757,102 (6.60%) | 0.586 $ | 1,006,373 | 0.821 $ | 1,750,729 | 0.451 $ |
| | iOS | 305,144 (0.73%) | 0.337 $ | 79,074 | 0.436 $ | 226,070 | 0.302 $ |
| | Unknown | 1,946,636 (4.66%) | 0.712 $ | 1,851,195 | 0.730 $ | 95,441 | 0.374 $ |
| **Ad Position** | Above the Fold | 23,127,357 (55.37%) | 0.657 $ | 21,027,882 | 0.689 $ | 2,099,475 | 0.337 $ |
| | Below the Fold | 18,640,606 (44.63%) | 0.592 $ | 14,487,566 | 0.693 $ | 4,153,040 | 0.242 $ |
| **Ad Format** | Large Banner Ad | 34,715 (0.08%) | 0.339 $ | 26,666 | 0.339 $ | 8,049 | 0.339 $ |
| | Billboard Ad | 94,480 (0.23%) | 8.109 $ | 93,801 | 8.122 $ | 679 | 6.373 $ |
| | Billboard Interstital Ad | 939,269 (2.25%) | 0.430 $ | 786,480 | 0.442 $ | 152,789 | 0.369 $ |
| | Extra Large Mobile Banner Ad | 2,986 (0.01%) | 0.610 $ | 2,956 | 0.612 $ | 30 | 0.430 $ |
| | Fullsize Mobile Ad | 2 (0.00%) | 10.224 $ | 2 | 10.224 $ | - | - |
| | Halfpage Ad | 67,463 (0.16%) | 5.653 $ | 67,416 | 5.654 $ | 47 | 4.907 $ |
| | Medium Rectangle Ad | 11,798,845 (28.25%) | 0.723 $ | 10,649,740 | 0.747 $ | 1,149,105 | 0.503 $ |
| | Mobile Banner Ad | 136,789 (0.33%) | 0.263 $ | 92,391 | 0.342 $ | 44,398 | 0.098 $ |
| | Mobile Leaderboard Ad | 3,596,930 (8.61%) | 0.113 $ | 1,018,680 | 0.156 $ | 2,578,250 | 0.096 $ |
| | Mobile Leaderboard (Wide) Ad | 458,020 (1.10%) | 0.135 $ | 429,671 | 0.138 $ | 28.349 | 0.092 $ |
| | Skyscraper Ad | 910,174 (2.18%) | 0.929 $ | 840,970 | 0.987 $ | 69,204 | 0.228 $ |
| | Standard Banner Ad | 14,984 (0.04%) | 1.391 $ | 14,969 | 1.391 $ | 15 | 1.097 $ |
| | Superwide Skyscraper Ad | 1,853,595 (4.44%) | 0.523 $ | 1,752,815 | 0.540 $ | 100,780 | 0.221 $ |
| | Takeover Ad | 73,643 (0.18%) | 0.339 $ | 59,979 | 0.340 $ | 14,664 | 0.339 $ |

| Variable | | All Ad Impressions | | Ad Impressions with User Tracking | | Ad Impressions without User Tracking | |
|---|---|---|---|---|---|---|---|
| | | Number of Ad Impressions (% of total) | Avg. Price (CPM) | Number of Ad Impressions | Avg. Price (CPM) | Number of Ad Impressions. | Avg. Price (CPM) |
| | Wallpaper Ad | 5,024,939 (12.03%) | 0.869 $ | 4,468,574 | 0.898 $ | 556,365 | 0.635 $ |
| | Wide Skyscraper Ad | 16,761,129 (40.13%) | 0.561 $ | 15,210,338 | 0.591 $ | 1,550,791 | 0.270 $ |
| **Week** | Week 1 | 21,327,958 (51.06%) | 0.608 $ | 18,260,710 | 0.665 $ | 3,067,248 | 0.270 $ |
| | Week 2 | 20,440,005 (48.94%) | 0.649 $ | 17,254,738 | 0.717 $ | 3,185,267 | 0.278 $ |
| **Weekday** | Monday | 6,436,688 (15.41%) | 0.615 $ | 5,652,574 | 0.660 $ | 784,114 | 0.289 $ |
| | Tuesday | 6,319,472 (15.13%) | 0.621 $ | 5,533,128 | 0.668 $ | 786,344 | 0.290 $ |
| | Wednesday | 5,832,770 (13.96%) | 0.626 $ | 5,036,253 | 0.681 $ | 796,517 | 0.282 $ |
| | Thursday | 5,980,117 (14.32%) | 0.667 $ | 5,245,727 | 0.717 $ | 734,390 | 0.313 $ |
| | Friday | 5,561,138 (13.31%) | 0.644 $ | 4,760,822 | 0.706 $ | 800,316 | 0.279 $ |
| | Saturday | 4,859,691 (11.63%) | 0.619 $ | 3,811,241 | 0.722 $ | 1,048,450 | 0.244 $ |
| | Sunday | 6,778,087 (16.23%) | 0.609 $ | 5,475,703 | 0.694 $ | 1,302,384 | 0.249 $ |
| **Hour of Day** | 0:00 | 284,473 (0.68%) | 0.709 $ | 211,851 | 0.862 $ | 72,622 | 0.262 $ |
| | 1:00 | 146,118 (0.35%) | 0.749 $ | 104,574 | 0.928 $ | 41,544 | 0.298 $ |
| | 2:00 | 88,070 (0.21%) | 0.876 $ | 60,914 | 1.119 $ | 27,156 | 0.331 $ |
| | 3:00 | 67,958 (0.16%) | 0.967 $ | 45,652 | 1.277 $ | 22,306 | 0.333 $ |
| | 4:00 | 80,729 (0.19%) | 0.953 $ | 56,416 | 1.209 $ | 24,313 | 0.360 $ |
| | 5:00 | 168,571 (0.40%) | 0.876 $ | 127,720 | 1.056 $ | 40,851 | 0.312 $ |
| | 6:00 | 457,851 (1.10%) | 0.795 $ | 372,372 | 0.911 $ | 85,479 | 0.290 $ |
| | 7:00 | 1,121,909 (2.69%) | 0.771 $ | 980,575 | 0.839 $ | 141,334 | 0.303 $ |
| | 8:00 | 1,720,997 (4.12%) | 0.699 $ | 1,534,779 | 0.746 $ | 186,218 | 0.313 $ |
| | 9:00 | 2,306,307 (5.52%) | 0.644 $ | 2,062,693 | 0.686 $ | 243,614 | 0.286 $ |
| | 10:00 | 2,537,895 (6.08%) | 0.626 $ | 2,268,315 | 0.667 $ | 269,580 | 0.280 $ |
| | 11:00 | 2,717,492 (6.51%) | 0.611 $ | 2,424,774 | 0.650 $ | 292,718 | 0.280 $ |
| | 12:00 | 2,781,084 (6.66%) | 0.618 $ | 2,450,380 | 0.665 $ | 330,704 | 0.276 $ |
| | 13:00 | 2,685,259 (6.43%) | 0.631 $ | 2,346,911 | 0.681 $ | 338,348 | 0.283 $ |
| | 14:00 | 2,753,561 (6.59%) | 0.631 $ | 2,391,380 | 0.684 $ | 362,181 | 0.282 $ |
| | 15:00 | 2,818,440 (6.75%) | 0.614 $ | 2,407,206 | 0.674 $ | 441,234 | 0.263 $ |
| | 16:00 | 2,828,678 (6.77%) | 0.605 $ | 2,392,470 | 0.668 $ | 436,208 | 0.260 $ |
| | 17:00 | 2,885,304 (6.91%) | 0.608 $ | 2,437,788 | 0.672 $ | 447,516 | 0.263 $ |
| | 18:00 | 2,730,669 (6.54%) | 0.622 $ | 2,340,221 | 0.678 $ | 390,448 | 0.286 $ |
| | 19:00 | 2,771,521 (6.64%) | 0.607 $ | 2,315,943 | 0.674 $ | 455,578 | 0.263 $ |
| | 20:00 | 2,597,226 (6.22%) | 0.594 $ | 2,125,461 | 0.668 $ | 471,765 | 0.261 $ |
| | 21:00 | 2,306,522 (5.52%) | 0.593 $ | 1,822,710 | 0.680 $ | 483,812 | 0.262 $ |
| | 22:00 | 1,862,366 (4.46%) | 0.586 $ | 1,423,897 | 0.686 $ | 438,469 | 0.259 $ |
| | 23:00 | 1,048,963 (2.51%) | 0.633 $ | 810,446 | 0.737 $ | 238,517 | 0.281 $ |

Notes: We did not include descriptive statistics for 1,344,474 user IDs, 16,500 city IDs, 111 publisher IDs, and 4,563 advertiser IDs.

*Table A4. Study 1a: Results Estimation of Probability of User Tracking Presence*

| **Dependent Variable: Log(Price)** | **Logit for AIPW** | **Probit for Heckman** |
|---|---|---|
| Log(Google Trends “http cookie”) | - | -0.004***<br>(0.001) |
| Device | Yes | Yes |
| OS | Yes | Yes |
| Browser | Yes | Yes |
| Time of Day, Weekday, Week | Yes | Yes |
| Publisher ID | Yes | Yes |
| AIC | 20,939,784 | 20,993,120 |
| N Ad Impressions | 41,767,963 | 41,767,963 |

*$p < 0.1$; **$p < 0.05$; ***$p < 0.01$.
Notes: AIPW = Augmented Inverse Probability Weighting. Robust standard errors are in parentheses.

*Table A5. Study 1a: Regression Results Robustness Checks*

| **Dependent Variable: Log(Price)** | **Linear Regression** (Full Data) | **Linear Regression** (98% Outlier Corrected Data) | **Linear Regression** (95% Outlier Corrected Data) | **AIPW** (Outcome Regression) | **Heckman** (Second Stage) |
|---|---|---|---|---|---|
| User Tracking (1/0) | 0.200*** | 0.194*** | 0.194*** | 0.201*** | 0.202*** |
| | (0.042) | (0.042) | (0.042) | (0.042) | (0.041) |
| Inverse Mills Ratio | | | | | 0.452*** |
| | | | | | (0.045) |
| Other User Data | | | | | |
| Device x Operating System × Browser | Yes | Yes | Yes | Yes | Yes |
| Time (Week, Weekday, Hour of Day) | Yes | Yes | Yes | Yes | Yes |
| Location of User (City) | Yes | Yes | Yes | Yes | Yes |
| Ad Slot Data | | | | | |
| Ad Position (Above Fold) | Yes | Yes | Yes | Yes | Yes |
| Ad Format | Yes | Yes | Yes | Yes | Yes |
| Publisher ID | Yes | Yes | Yes | Yes | Yes |
| Control | | | | | |
| Advertiser ID | Yes | Yes | Yes | Yes | Yes |
| Adj. R2 | 0.654 | 0.647 | 0.654 | 0.654 | 0.654 |
| N Ad Impressions | 41,767,963 | 40,937,021[A] | 39,757,090[A] | 41,430,997[B] | 41,767,963 |

*$p$ < 0.1; **$p$ < 0.05; ***$p$ < 0.01. Notes: Robust standard errors are in parentheses.
AIPW: Augmented Inverse Probability Weighting

[A] To account for the potential effect of outliers, we exclude all ad impressions with a price > $2.98 (= 830,942 ad impressions, 1.98%) in the 98% outlier corrected data set, and all ad impressions with a price > 1.55 (= 2,010,873, 4.8%) in the 95% outlier corrected data set.

[B] In line with previous research, we exclude 336,966 (0.80%) ad impressions for the AIPW estimate with a very high ($\hat{\pi}_i > 0.999$) or very low ($\hat{\pi}_i < 0.001$) treatment probability because they could yield very high or very low weights.

*Table A6. Study 1a: Standardized Mean Differences in Covariates Between Ad Impressions With User Tracking and Without User Tracking Before and After Weighting*

| | | Unweighted | | | Weighted | | |
|---|---|---|---|---|---|---|---|
| **Variable** | | **Mean (SD) without User Tracking** | **Mean (SD) with User Tracking** | **SMD** | **Mean (SD) without User Tracking** | **Mean (SD) with User Tracking** | **SMD** |
| **Device** | Desktop | 0.42 (0.49) | 0.91 (0.29) | 1.198 | 0.82 (0.38) | 0.83 (0.37) | 0.024 |
| | Smartphone | 0.24 (0.43) | 0.03 (0.17) | 0.64 | 0.06 (0.23) | 0.06 (0.24) | 0.029 |
| | Tablet | 0.14 (0.35) | 0.02 (0.16) | 0.434 | 0.06 (0.25) | 0.04 (0.20) | 0.09 |
| | Unknown | 0.20 (0.40) | 0.04 (0.19) | 0.512 | 0.06 (0.23) | 0.06 (0.24) | 0.02 |
| **Operating System** | Android | 0.40 (0.49) | 0.06 (0.25) | 0.869 | 0.11 (0.31) | 0.12 (0.32) | 0.031 |
| | Apple Macintosh | 0.15 (0.36) | 0.02 (0.14) | 0.488 | 0.04 (0.19) | 0.04 (0.19) | 0.005 |
| | Apple iOS | 0.17 (0.37) | 0.01 (0.12) | 0.552 | 0.04 (0.19) | 0.04 (0.19) | 0.001 |
| | BlackBerry OS | 0.00 (0.01) | 0.00 (0.02) | 0.024 | 0.00 (0.02) | 0.00 (0.02) | 0 |
| | Linux | 0.00 (0.04) | 0.00 (0.05) | 0.029 | 0.00 (0.05) | 0.00 (0.05) | 0.001 |
| | Microsoft Windows | 0.27 (0.44) | 0.89 (0.32) | 1.599 | 0.80 (0.40) | 0.79 (0.40) | 0.027 |
| | Symbian OS | 0.00 (0.00) | 0.00 (0.00) | 0.005 | 0.00 (0.00) | 0.00 (0.00) | 0 |
| | Unknown | 0.01 (0.10) | 0.01 (0.11) | 0.032 | 0.01 (0.11) | 0.01 (0.11) | 0 |
| **Browser** | Android | 0.00 (0.01) | 0.01 (0.11) | 0.156 | 0.00 (0.05) | 0.01 (0.10) | 0.108 |
| | Chrome | 0.45 (0.50) | 0.19 (0.39) | 0.563 | 0.23 (0.42) | 0.23 (0.42) | 0.004 |
| | Firefox | 0.17 (0.38) | 0.45 (0.50) | 0.629 | 0.40 (0.49) | 0.41 (0.49) | 0.012 |
| | Internet Explorer | 0.04 (0.20) | 0.25 (0.43) | 0.623 | 0.24 (0.42) | 0.22 (0.41) | 0.037 |
| | Opera | 0.01 (0.10) | 0.01 (0.10) | 0.009 | 0.01 (0.10) | 0.01 (0.10) | 0.002 |
| | Safari | 0.28 (0.45) | 0.03 (0.17) | 0.743 | 0.07 (0.25) | 0.07 (0.25) | 0.003 |
| | iOS | 0.04 (0.19) | 0.00 (0.05) | 0.249 | 0.01 (0.08) | 0.01 (0.09) | 0.005 |
| | Unknown | 0.02 (0.12) | 0.05 (0.22) | 0.207 | 0.05 (0.22) | 0.05 (0.21) | 0.01 |
| **Ad Position** | Above the Fold | 0.34 (0.47) | 0.59 (0.49) | 0.536 | 0.55 (0.50) | 0.56 (0.50) | 0.004 |
| **Ad Format** | Large Banner Ad | 0.00 (0.04) | 0.00 (0.03) | 0.018 | 0.00 (0.03) | 0.00 (0.03) | 0.003 |
| | Billboard Ad | 0.00 (0.01) | 0.00 (0.05) | 0.068 | 0.00 (0.01) | 0.00 (0.05) | 0.063 |
| | Billboard Interstital Ad | 0.02 (0.15) | 0.02 (0.15) | 0.017 | 0.02 (0.15) | 0.02 (0.15) | 0.003 |
| | Extra Large Mobile Banner Ad | 0.00<br>(0.00) | 0.00<br>(0.01) | 0.011 | 0.00<br>(0.00) | 0.00<br>(0.01) | 0.013 |
| | Fullsize Mobile Ad | 0.00<br>(0.00) | 0.00<br>(0.00) | 0.001 | 0.00<br>(0.00) | 0.00<br>(0.00) | 0 |
| | Halfpage Ad | 0.00<br>(0.00) | 0.00<br>(0.04) | 0.061 | 0.00<br>(0.00) | 0.00<br>(0.04) | 0.06 |
| | Medium Rectangle Ad | 0.18<br>(0.39) | 0.30<br>(0.46) | 0.274 | 0.30<br>(0.46) | 0.28<br>(0.45) | 0.028 |
| | Mobile Banner Ad | 0.01<br>(0.08) | 0.00<br>(0.05) | 0.071 | 0.00<br>(0.06) | 0.00<br>(0.05) | 0.007 |
| | Mobile Leaderboard Ad | 0.41<br>(0.49) | 0.03<br>(0.16) | 1.048 | 0.09<br>(0.28) | 0.06<br>(0.25) | 0.079 |
| | Mobile Leaderboard (Wide) Ad | 0.00<br>(0.07) | 0.01<br>(0.11) | 0.082 | 0.00<br>(0.04) | 0.03<br>(0.17) | 0.237 |
| | Skyscraper Ad | 0.01<br>(0.10) | 0.02<br>(0.15) | 0.098 | 0.02<br>(0.14) | 0.02<br>(0.15) | 0.024 |
| | Standard Banner Ad | 0.00<br>(0.00) | 0.00<br>(0.02) | 0.028 | 0.00<br>(0.00) | 0.00<br>(0.02) | 0.029 |
| | Superwide Skyscraper Ad | 0.02<br>(0.13) | 0.05<br>(0.22) | 0.188 | 0.03<br>(0.16) | 0.05<br>(0.21) | 0.1 |
| | Takeover Ad | 0.00<br>(0.05) | 0.00<br>(0.04) | 0.012 | 0.00<br>(0.04) | 0.00<br>(0.04) | 0.004 |
| | Wallpaper Ad | 0.09<br>(0.28) | 0.13<br>(0.33) | 0.118 | 0.13<br>(0.33) | 0.12<br>(0.33) | 0.017 |
| | Wide Skyscraper Ad | 0.25<br>(0.43) | 0.43<br>(0.50) | 0.392 | 0.42<br>(0.49) | 0.40<br>(0.49) | 0.034 |

*Table A7. Study 1a: Price Predictions (Potential Outcomes) for Value of Ad Obtrusiveness and Ad Visibility*

| | **Price Predictions and Treatment Effects** | | | |
|---|---|---|---|---|
| | **High Visibility** | **Low Visibility** | **High Obtrusiveness** | **Low Obtrusiveness** |
| **E(Price\| User Tracking =1)** | 0.680 | 0.601 | 0.766 | 0.636 |
| **E(Price\| User Tracking =0)** | 0.480 | 0.475 | 0.730 | 0.458 |
| **ATE (in US$)** | **-0.200** | **-0.126** | **-0.036** | **-0.178** |
| **ATE (%) / Relative Price Difference** | **-29.4%** | **-20.9%** | **-5.7%** | **-27.9%** |
| N Ad Impressions | 41,767,963 | | 41,767,963 | |

Notes: ATE = Average Treatment Effect. High Obtrusiveness = Large Ads = 1. High Visibility = Ad Above Fold = 1. Correlation ad above the fold and large ad = 0.047. Number of publishers with ads above fold: 52. Number of publishers with large ads = 93. Number of unique ad formats per publisher: Mean = 5.21, Min = 1, Max = 9, SD = 1.83, n = 111

*Table A8. Study 1a: Price Predictions (Potential Outcomes) Premium, Thematic Content, and Size*

| | Price Predictions and Treatment Effects | | | | | |
|---|---|---|---|---|---|---|
| | **Premium Publisher** | **Non-Premium Publisher** | **Thematic-focused Publisher** | **Thematic-broad Publisher** | **Large Publisher** | **Small Publisher** |
| **E(Price\| User Tracking =1)** | 0.725 | 0.621 | 1.060 | 0.644 | 0.647 | 0.722 |
| **E(Price\| User Tracking =0)** | 0.578 | 0.444 | 1.050 | 0.473 | 0.477 | **0.630** |
| **ATE (in US$)** | **-0.147** | **-0.177** | **0.010** | **-0.171** | **-0.17** | **-0.092** |
| **ATE (%) / Relative Price Difference** | **-20.3%** | **-28.5%** | **-0.9%** | **-26.6%** | **-26.3%** | **-12.7%** |
| N Ad Impressions | 41,767,963 | | 41,767,963 | | 41,767,963 | |

Notes: ATE = Average Treatment Effect. Correlation of premium publisher and thematic-focused publisher (-0.170), correlation of premium publisher and large publisher (0.390), correlation of large publisher and thematic-focused publisher (-0.320). Number of premium publishers = 24 (22%). Number of thematic-focused publishers = 35 (32%). Number of large publishers = 55 (50%).

*Table A9. Study 1a: Coefficients for Ad Frequency Groups and Ad Recency Groups from Model 4.1*

| **Coefficients for Ad Frequency Groups and Ad Recency Groups** | **Model 4.1** (see Table 7 for remaining coefficients) |
|---|---|
| Advertiser frequency = 01 | -0.088*** (0.007) |
| Advertiser frequency = 02 | -0.094*** (0.009) |
| Advertiser frequency = 03 | -0.097*** (0.010) |
| Advertiser frequency = 04 | -0.112*** (0.011) |
| Advertiser frequency = 05 | -0.107*** (0.011) |
| Advertiser frequency = 06 | -0.107*** (0.011) |
| Advertiser frequency = 07 | -0.104*** (0.012) |
| Advertiser frequency = 08 | -0.1208*** (0.012) |
| Advertiser frequency = 09 | -0.118*** (0.012) |
| Advertiser frequency = 10 | -0.116*** (0.012) |
| Advertiser frequency = 11 - 20 | -0.211*** (0.013) |
| Advertiser frequency = 21 - 30 | -0.205*** (0.014) |
| Advertiser frequency = 31 - 40 | -0.193*** (0.014) |
| Advertiser frequency = 41 - 50 | -0.183*** (0.014) |
| Advertiser frequency = 51 - 60 | -0.169*** (0.014) |
| Advertiser frequency = 61 - 70 | -0.165*** (0.014) |
| Advertiser frequency = 71 - 80 | -0.159*** (0.014) |
| Advertiser frequency = 81 - 90 | -0.156*** (0.014) |
| Advertiser frequency = 91 - 100 | -0.154*** (0.014) |
| Advertiser frequency > 100 | -0.143*** (0.018) |
| | |
| Advertiser recency = 01 | -0.143*** (0.002) |
| Advertiser recency = 02 | -0.175*** (0.003) |
| Advertiser recency = 03 | -0.191*** (0.004) |
| Advertiser recency = 04 | -0.199*** (0.004) |
| Advertiser recency = 05 | -0.174*** (0.004) |
| Advertiser recency = 06 | -0.198*** (0.004) |
| Advertiser recency = 07 | -0.203*** (0.003) |
| Advertiser recency = 08 | -0.199*** (0.003) |
| Advertiser recency = 09 | -0.194*** (0.003) |
| Advertiser recency = 10 | -0.185*** (0.003) |
| Advertiser recency = 11 - 20 | -0.138*** (0.003) |
| Advertiser recency = 21 - 30 | -0.060*** (0.005) |
| Advertiser recency = 31 - 40 | -0.046*** (0.006) |
| Advertiser recency = 41 - 50 | -0.043*** (0.006) |
| Advertiser recency = 51 - 60 | -0.045*** (0.006) |
| Advertiser recency = 61 - 70 | -0.050*** (0.006) |
| Advertiser recency = 71 - 80 | -0.084*** (0.005) |
| Advertiser recency = 81 - 90 | -0.095*** (0.004) |
| Advertiser recency = 91 - 100 | -0.106*** (0.004) |
| Advertiser recency > 100 | -0.005 (0.006) |

*$p < 0.1$; **$p < 0.05$; ***$p < 0.01$.
Notes: Robust standard errors are in parentheses. Baseline: Ad frequency = 0, and ad recency = 0.

*Table A10. Study 1a: Price Differences for Value Added by Browsing History Across Publisher Type and Publisher Content Categories*

| Publisher Type / Publisher Content Category | Price Differences (In US$) | | | | | | | |
|---|---|---|---|---|---|---|---|---|
| | Mean | Sd. | Quantiles | | | | | |
| | | | 0% | 50% | 75% | 90% | 98% | 100% |
| *Panel A: Value Added by Browsing History by Publisher Type* | | | | | | | | |
| All Publishers | 0.002 | 0.007 | -2.391 | 0.002 | 0.003 | 0.004 | 0.009 | 2.131 |
| Premium | 0.004 | 0.005 | -1.062 | 0.003 | 0.004 | 0.007 | 0.015 | 2.131 |
| Non-Premium | 0.002 | 0.007 | -2.392 | 0.002 | 0.002 | 0.004 | 0.007 | 1.223 |
| Thematic-Focused | -0.002 | 0.046 | -2.392 | 0.003 | 0.004 | 0.007 | 0.018 | 1.223 |
| Thematic-Broad | 0.002 | 0.005 | -1.719 | 0.002 | 0.003 | 0.004 | 0.009 | 2.131 |
| Large Publisher | 0.002 | 0.006 | -2.392 | 0.002 | 0.003 | 0.004 | 0.009 | 2.131 |
| Small Publisher | -0.001 | 0.038 | -1.295 | 0.002 | 0.004 | 0.007 | 0.022 | 0.380 |
| *Panel B: Value Added by Browsing History by Publisher Content Category* | | | | | | | | |
| Health | 0.003 | 0.005 | -0.205 | 0.003 | 0.005 | 0.006 | 0.011 | 0.797 |
| News & Information | 0.002 | 0.003 | -1.719 | 0.002 | 0.002 | 0.004 | 0.007 | 1.155 |
| Sports | 0.002 | 0.003 | -0.515 | 0.002 | 0.003 | 0.005 | 0.008 | 0.162 |
| Lifestyle & Shopping | 0.007 | 0.019 | -0.784 | 0.005 | 0.008 | 0.014 | 0.028 | 1.675 |
| Women's Interests | 0.005 | 0.008 | -0.363 | 0.004 | 0.006 | 0.010 | 0.025 | 0.336 |
| Student's Interests | 0.008 | 0.014 | -0.033 | 0.004 | 0.007 | 0.016 | 0.052 | 0.642 |
| Games | -0.019 | 0.067 | -2.392 | 0.001 | 0.002 | 0.004 | 0.007 | 0.080 |
| Cars | 0.004 | 0.007 | -0.369 | 0.003 | 0.004 | 0.007 | 0.181 | 1.207 |
| Travel | 0.004 | 0.005 | -0.030 | 0.003 | 0.005 | 0.007 | 0.014 | 0.414 |
| Entertainment | 0.005 | 0.008 | -1.062 | 0.003 | 0.005 | 0.008 | 0.024 | 2.131 |
| Computers & Technology | 0.005 | 0.006 | -0.710 | 0.004 | 0.006 | 0.009 | 0.019 | 0.340 |
| Dating | -0.017 | 0.032 | -1.065 | -0.005 | -0.000 | -.001 | 0.002 | 0.023 |
| Money, Finance & Real Estate | 0.005 | 0.005 | -0.181 | 0.004 | 0.006 | 0.009 | 0.019 | 0.349 |

*Table A11. Study 2: Descriptive Statistics*

| Variable | | All Ad Impressions | | Ad Impressions with User Tracking | | Ad Impressions without User Tracking | |
|---|---|---|---|---|---|---|---|
| | | Number of Ad Impressions (% of total) | Avg. Price (CPM) | Number of Ad Impressions | Avg. Price (CPM) | Number of Ad Impressions. | Avg. Price (CPM) |
| Panel A: Europe *(N = 10,433,115)* | | | | | | | |
| **Operating System** | Android | 4,258,183 (40.81%) | 7.732 $ | 3,800,863 | 8.217 $ | 457,320 | 3.706 $ |
| | iOS | 6,174,932 (59.19%) | 6.131 $ | 1,115,062 | 6.452 $ | 5,059,870 | 6.054 $ |
| **Ad Format** | Banner | 3,345,821 (32,07% | 0.739 $ | 1,047,711 | 1.009 $ | 2,298,110 | 0.616 $ |
| | Interstitial | 5,511,119 (52,82%) | 8.333 $ | 3,108,077 | 8.245 $ | 2,403,042 | 8.447 $ |
| | Rewarded | 1,576,175 (15,1%) | 14.186 $ | 760,137 | 15.451 $ | 816,038 | 13.008 $ |
| *Panel B: United States (N = 207,961,593)* | | | | | | | |
| **Operating System** | Android | 52,531,708 (25.26%) | 8.854 $ | 47,972,052 | 9.005 $ | 4,559,656 | 7.270 $ |
| | iOS | 155,429,885 (74.74%) | 5.171 $ | 26,511,907 | 7.804 $ | 128,917,978 | 4.630 $ |
| **Ad Format** | Banner | 135,077,447 | 0.831 $ | 41,968,333 | 1.112 $ | 93,109,114 | 0.705 $ |
| | Interstitial | 53,752,442 | 13.746 $ | 24,776,376 | 16.127 $ | 28,976,066 | 11.709 $ |
| | Rewarded | 19,131,704 | 21.836 $ | 7,739,250 | 24.892 $ | 11,392,454 | 19.761 $ |

*Table A12. Study 2: Distribution of log(Ad Impression Prices) with and without User Tracking*

| (CPM) Price in US$ | Mean | Standard Deviation | Median | Min | q95 | q98 | Max | N ( = Ad Impresions) |
|---|---|---|---|---|---|---|---|---|
| | | | | *Panel A: Europe (N = 10,433,115)* | | | | |
| With User Tracking | 1.508 | 1.307 | 1.926 | -3.222 | 2.926 | 3.359 | 4.460 | 4,915,925 |
| Without User Tracking | 0.647 | 1.852 | 1.462 | -3.396 | 2.983 | 3.367 | 4.792 | 5,517,190 |
| | | | | *Panel B: United States (N = 207,961,593)* | | | | |
| With User Tracking | 0.954 | 1.619 | 0.438 | -2.443 | 3.576 | 3.914 | 5.414 | 74,483,959 |
| Without User Tracking | -0.098 | 1.834 | -0.674 | -2.831 | 3.088 | 3.603 | 5.414 | 133,477,634 |

Notes: $N_{Total}$ = 218,394,708; CPM Price = price for 1,000 ad impressions, min = minimum, max = maximum, q95 = 95% quantile, q98 = 98% quantile.

*Table A13. Study 2: Price Predictions (Potential Outcomes) for Value of Ad Obtrusiveness*

**Price Predictions and Treatment Effects**

| | **EU** | | **US** | |
|---|---|---|---|---|
| | **High Obtrusiveness** | **Low Obtrusiveness** | **High Obtrusiveness** | **Low Obtrusiveness** |
| **E(Price\| User Tracking =1)** | 7.38 $ | 0.756 | 7.40 | 1.05 |
| **E(Price\| User Tracking =0)** | 6.96 $ | 0.366 | 5.31 | 0.452 |
| **ATE (in US$)** | **-0.42** | **-0.39** | **-2.09** | **-0.598** |
| **ATE (%) / Relative Price Difference** | **-6.0%** | **-51.58%** | **-28.24%** | **-56.96%** |
| N Publisher-Instances / N Publishers / N Ad Impressions | 3,412 / 1,225 / 10,433,115 | | 28,478 / 9,301 / 207,961,593 | |

Notes: ATE = Average Treatment Effect.

*Figure B1. Study 1a: Kernel Density Plots of Raw and Logarithmized Ad Impression Prices (in US$) by Presence of User Tracking (N = 41,767,963)*

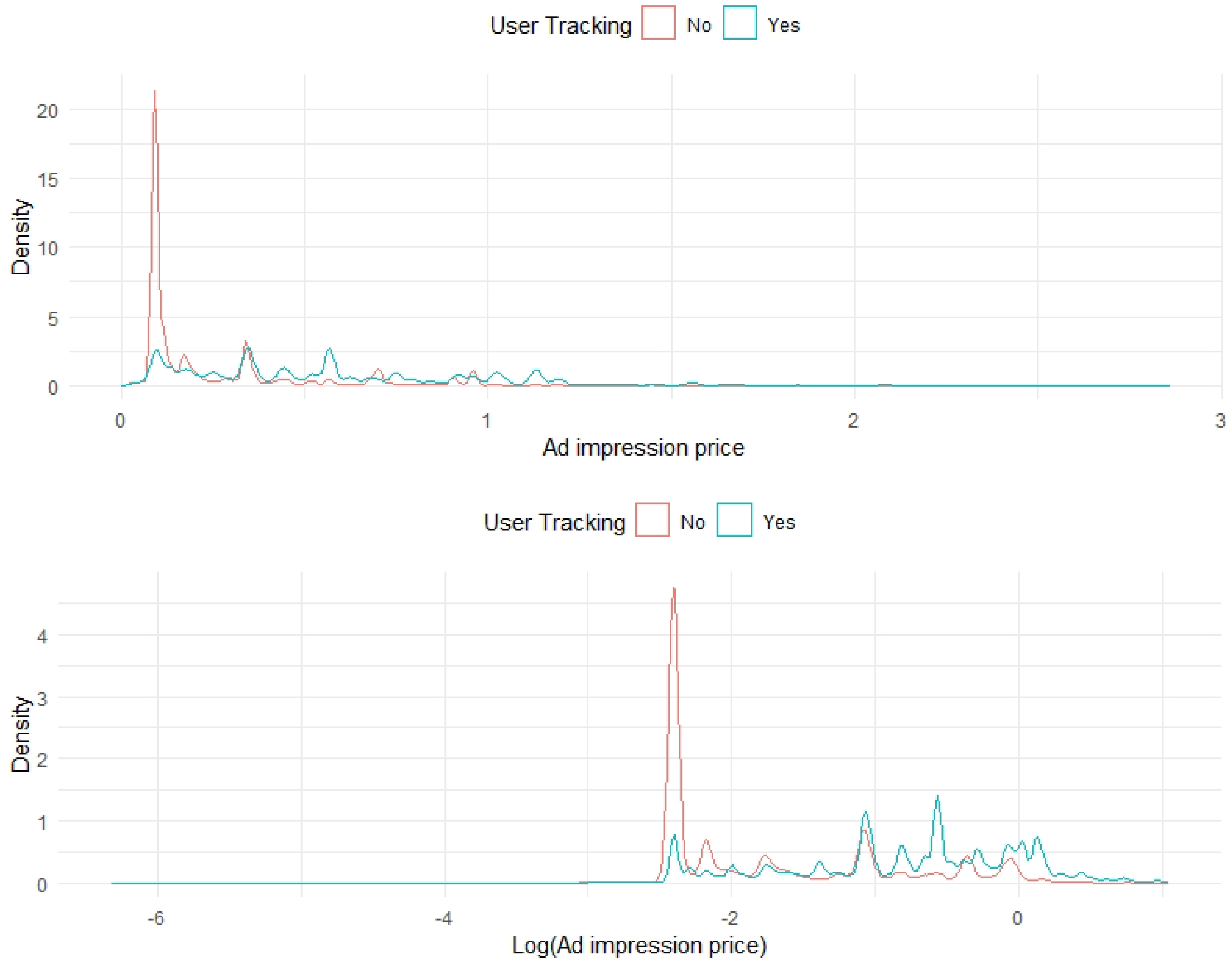

*Figure B2. Study 1a: Standardized Mean Differences in Covariates Between Ad Impressions with and without User Tracking Before and After Weighting*

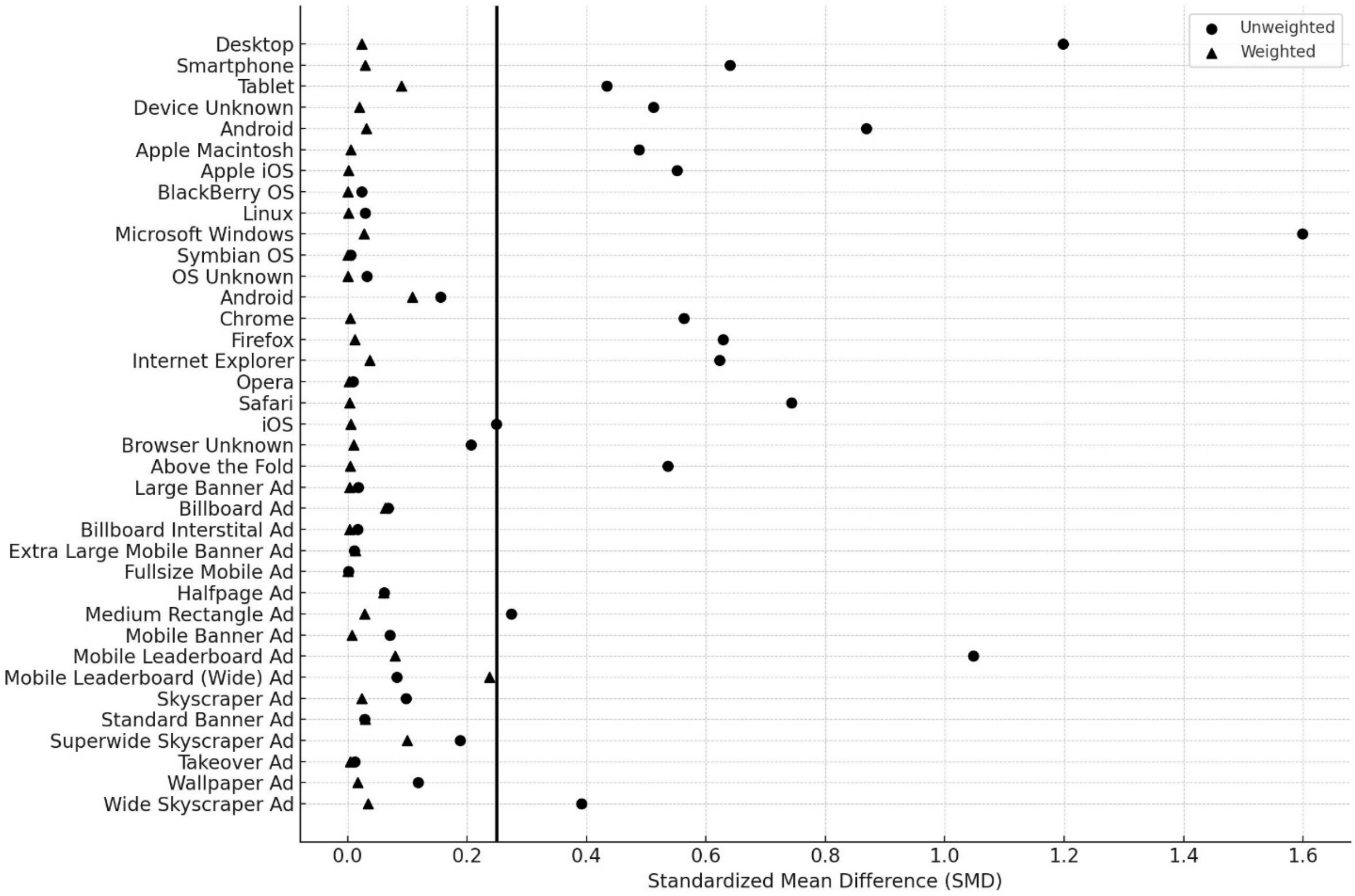

*Figure B3. Study 1a: Google Search Volume during the Observation Period (April 2016)*

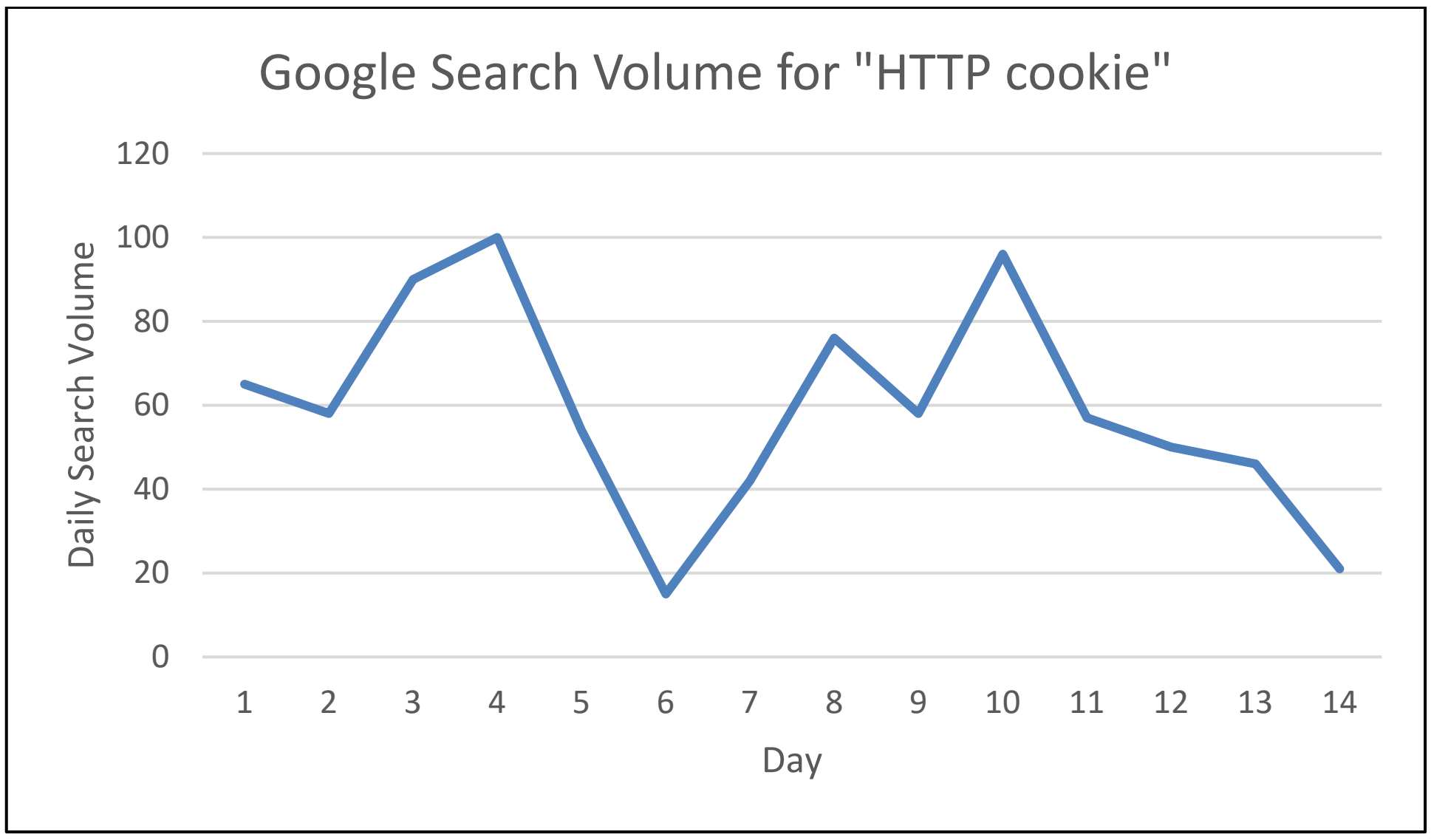

*Figure B4. Study 2: Kernel Density Plots of Ad Impression Prices (in US$) by Presence of User Tracking*

*Panel A: Europe (N = 10,433,115)*

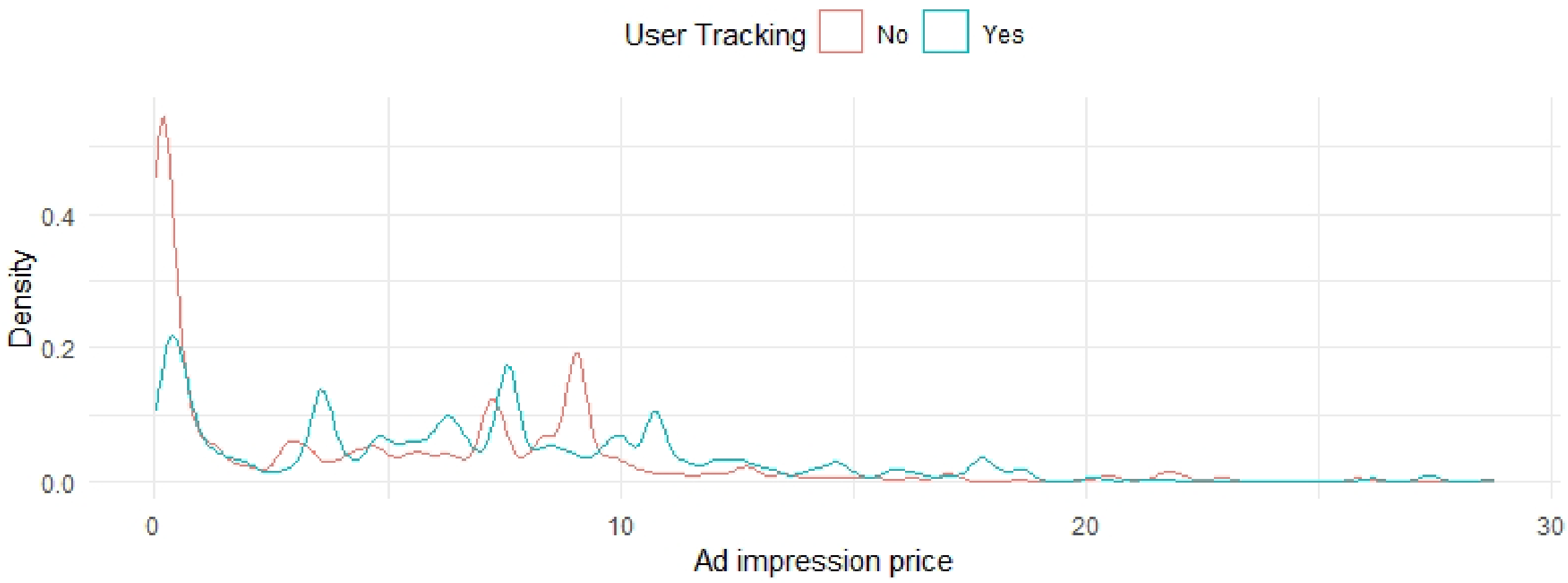

*Panel B: United States (N = 207,961,593)*

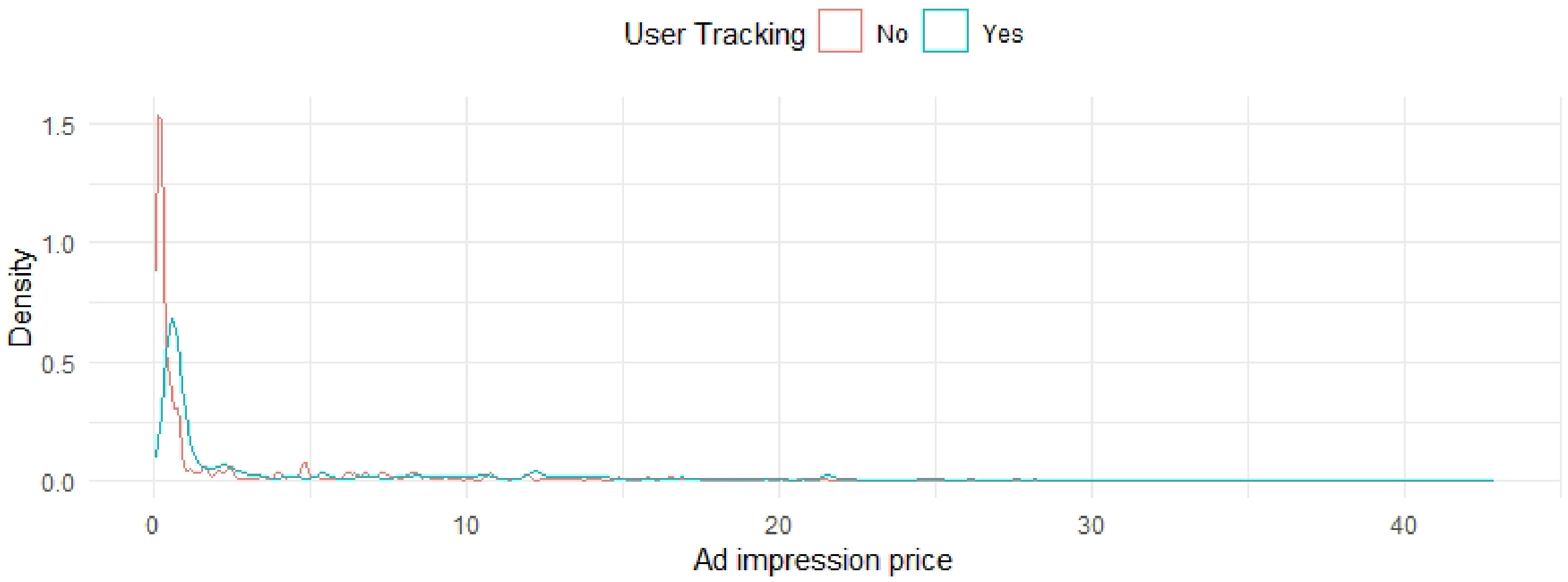

*Figure B5. Study 2: Kernel Density Plots of log(Ad Impression Prices (in US$)) by Presence of User Tracking*

*Panel A: Europe (N = 10,433,115)*

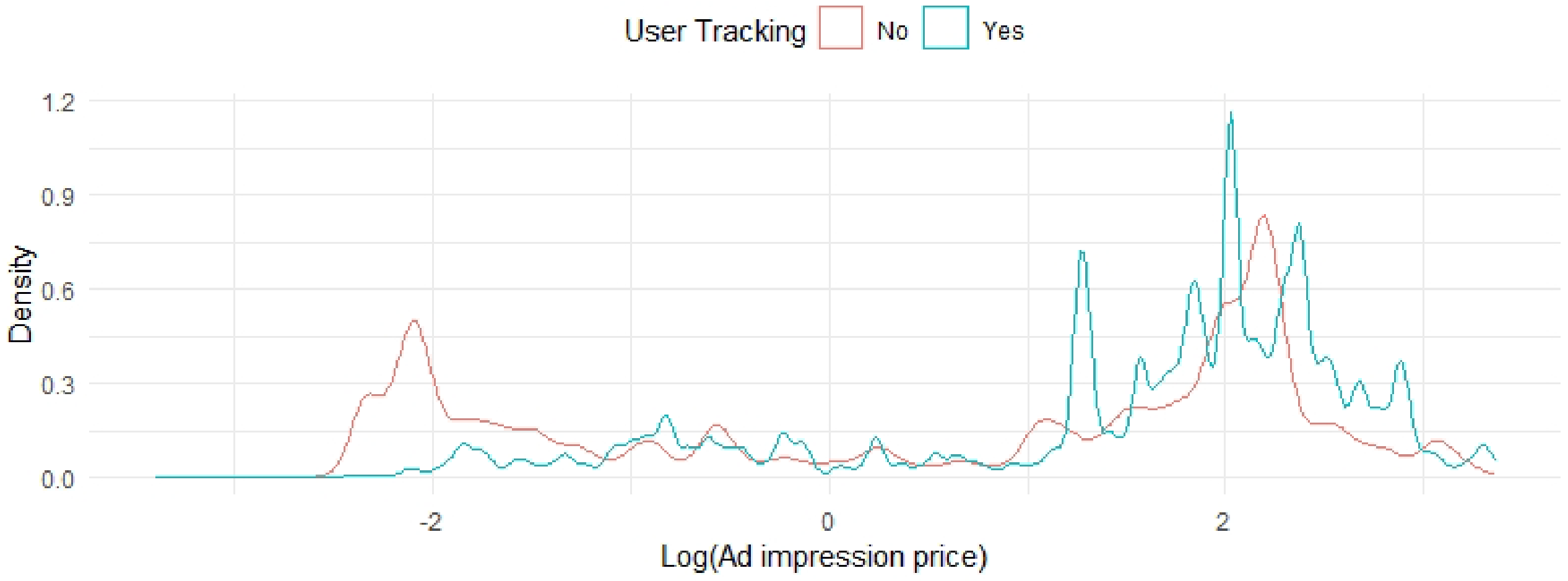

*Panel B: United States (N = 207,961,593)*

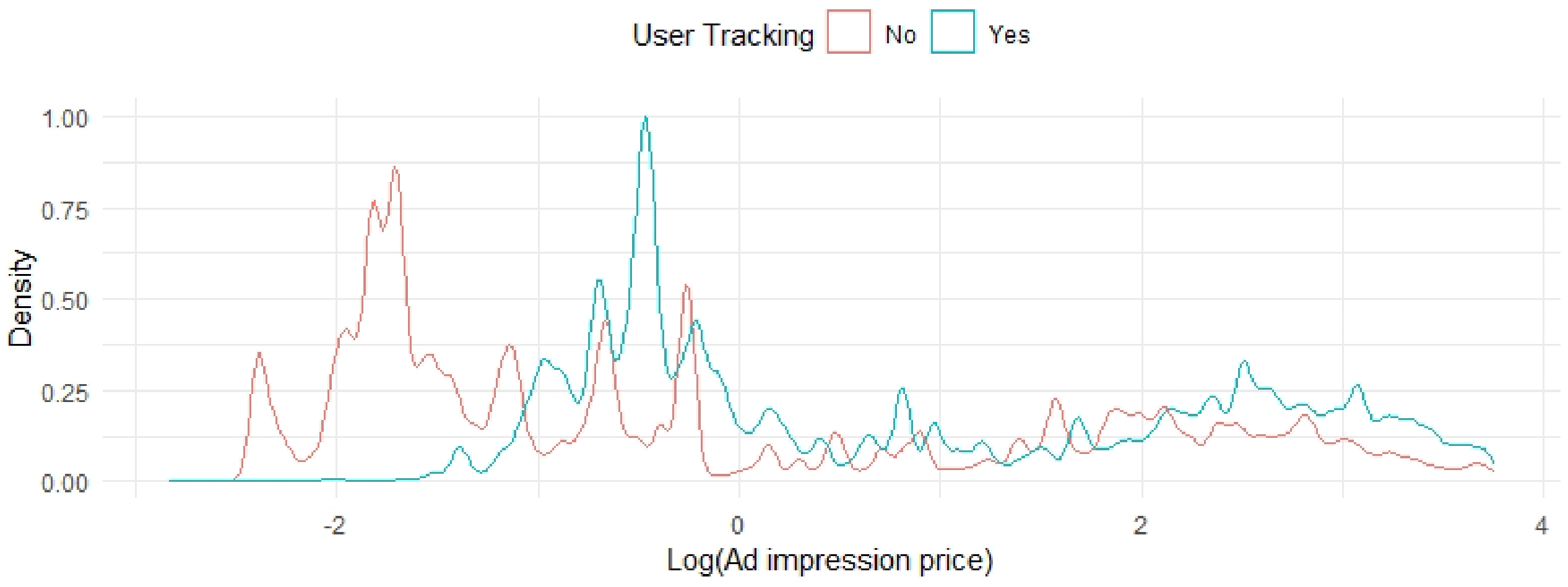